\documentclass[a4paper,twocolumn,11pt,accepted=2020-06-03]{quantumarticle}
\pdfoutput=1
\usepackage[utf8]{inputenc}
\usepackage[english]{babel}
\usepackage[T1]{fontenc}
\usepackage{amsmath}
\usepackage{hyperref}
\usepackage{ulem}

\usepackage{tikz}
\usepackage{lipsum}

\usepackage{color} 
\usepackage{longtable}
\usepackage{subfigure}
\usepackage{amsfonts}

\def\ket#1{|{#1}\rangle}

\def\e{\mathrm{e}}
\def\ii{\mathrm{i}}
\def\d{\mathrm{d}}

\newcommand{\lyxdot}{.}

\begin{document}

\title{Real Time Dynamics and Confinement in the $\mathbb{Z}_{n}$ Schwinger-Weyl lattice model for 1+1 QED}

\author{Giuseppe Magnifico}
\orcid{0000-0002-7280-445X}
\affiliation{Dipartimento di Fisica e Astronomia dell'Universit\`a di Bologna, I-40127 Bologna, Italy}
\affiliation{INFN, Sezione di Bologna, I-40127 Bologna, Italy}

\author{Marcello Dalmonte}
\orcid{0000-0001-5338-4181}
\affiliation{Abdus Salam ICTP, Strada Costiera 11, I-34151 Trieste, Italy}
\affiliation{SISSA, Via Bonomea 265, I-34136 Trieste, Italy}

\author{Paolo Facchi}
\orcid{0000-0001-9152-6515}
\affiliation{Dipartimento di Fisica and MECENAS, Universit\`{a} di Bari, I-70126 Bari, Italy}
\affiliation{INFN, Sezione di Bari, I-70126 Bari, Italy}

\author{Saverio Pascazio}
\orcid{0000-0002-7214-5685}
\affiliation{Dipartimento di Fisica and MECENAS, Universit\`{a} di Bari, I-70126 Bari, Italy}
\affiliation{INFN, Sezione di Bari, I-70126 Bari, Italy}
\affiliation{Istituto Nazionale di Ottica (INO-CNR), I-50125 Firenze, Italy}

\author{\\ Francesco V. Pepe}
\orcid{0000-0002-7407-063X}
\affiliation{Dipartimento di Fisica and MECENAS, Universit\`{a} di Bari, I-70126 Bari, Italy}
\affiliation{INFN, Sezione di Bari, I-70126 Bari, Italy}

\author{Elisa Ercolessi}
\orcid{0000-0002-6801-5976}
\affiliation{Dipartimento di Fisica e Astronomia dell'Universit\`a di Bologna, I-40127 Bologna, Italy}
\affiliation{INFN, Sezione di Bologna, I-40127 Bologna, Italy}

\begin{abstract}
We study the out-of-equilibrium properties of $1+1$ dimensional quantum electrodynamics (QED), discretized via the staggered-fermion Schwinger model with an Abelian $\mathbb{Z}_{n}$ gauge group. We look at two relevant phenomena: first, we analyze the stability of the Dirac vacuum with respect to particle/antiparticle pair production, both spontaneous and induced by an external electric field; then, we examine the string breaking mechanism. We observe a strong effect of confinement, which acts by suppressing both spontaneous pair production and string breaking into quark/antiquark pairs, indicating that the system dynamics displays a number of out-of-equilibrium features.
\end{abstract}

\maketitle

\section{Introduction}

Gauge fields coupled to matter are at the heart of the microscopical description of nature: they represent the key ingredient to describe the physics of fundamental particles and interactions and appear in a variety of problems of condensed matter physics, such as superconductivity or the Quantum Hall Effect~\cite{Kleinert,Fradkin}. Quantum field theory has developed by tackling more and more complicate issues, and by now we have not only a comprehensive description of the low energy sector of the model, but also techniques to study strong coupling, non-perturbative and topological effects. 
Numerical simulations, especially those based on Monte Carlo techniques~\cite{Creutz}, have been able to confirm many phenomena predicted with analytical methods and shed light on several non-trivial aspects. The possibility to describe  a given field theory through a lattice theory is an important problem~\cite{rothe1992lattice,montvay1997quantum}, which has been investigated by many physicists since the 70's~\cite{wilsonlgt,kogut1975hamiltonian,susskind1977lattice,kogut1979introduction}. 

It is clear that, from a theoretical point of view, it is important to determine how effectively a lattice model is able to reproduce the salient features of a given quantum gauge theory. Very often this offers a formidable computational problem, but recently techniques developed from quantum information insights, such as tensor networks~\cite{DMRG1,MPS}, have been used to tackle these questions, thanks to their intrinsic ability to restrict the dynamics to relevant subspaces of the Hilbert space, exploiting the entanglement of the states that contribute to the dynamics.
From an experimental point of view, recent advances in the control of ultra-cold atomic systems have paved the way to a whole new way of studying fundamental theories in the spirit of quantum simulations, as suggested by Feynman~\cite{Feyn}. At present, there are many proposals in the literature to use ultra-cold atoms or  trapped ions in optical lattices to simulate many-body Hamiltonians~\cite{bdz,qsim1,qsim2,qsim3,qsim4}, as well as Abelian and non-Abelian lattice gauge theories~\cite{simul1,simul4,simul3,simul5,simul6,simul7,simul9,simul11,zoharreview}. These ideas led to the first experimental realization of a quantum simulation of 1+1 dimensional quantum electrodynamics (QED) with four trapped ions~\cite{martinez2016}. On the other hand, the implementation of classical ``synthetic'' fields has been proposed and realized in a number of experiments~\cite{cmr,ytterbium,mpi_sun,fallani,lcd}. 

In general, an approach based on quantum simulators opens the possibility to study in a more efficient way problems that have always been difficult to tackle, including non-perturbative effects as well as genuine dynamical behavior. 
These phenomena are necessary to understand the essence of Abelian and non-Abelian quantum field theories in applications to particle physics. Two examples on which we will elaborate in this article are vacuum stability in QED with respect to particle/antiparticle pair production~\cite{Sch51} and confinement in quantum chromodynamics (QCD)~\cite{Wilc}, together with the associated string breaking mechanism. 
In recent years, much attention has also been devoted to the interplay between correlations and dynamics in statistical mechanics and many-body systems in the quantum realm, which may exhibit unusual thermalization and equilibration properties. It has been stressed that, in a closed quantum system, non-equilibrium dynamics might display behaviors that cannot be understood by means of the standard theory of ergodic or integrable models and the corresponding (Gibbs ensemble) thermalization hypothesis. This is the case of the so-called many-body localization phenomenon (for reviews of this topic see, for example,~\cite{Nand,Alet}) or of the recently discussed quantum scars~\cite{scar1,scar2}. The interest in these systems and phenomena has fostered the development of novel paradigms.

In the spirit of such research developments, in this article we shall tackle the problem of the out-of-equilibrium real-time dynamics of 1+1 dimensional QED, which represents the simplest model of fermionic matter coupled to (Abelian) gauge fields. In spite of its simplicity, the model shares many phenomenological features with more complicated theories, such as QCD, thus representing an interesting instance to test quantum field theory and quantum simulations. The model has been widely investigated since the 70's \cite{wilsonlgt,kogut1975hamiltonian,susskind1977lattice,kogut1979introduction} and has been recently proposed as a possible viable experimental option in the not-too-distant future~\cite{simul10,KCB,NEFMPP,simul12,simone}. Indeed, a first experiment reproducing such a model with few qubits in a ion trap has already been reported~\cite{martinez2016} and new ideas have been proposed for possible experimental realizations with Rydberg atoms~\cite{surace,simone}.

The first problem to face when trying to encode a lattice-gauge theory in a quantum simulation with cold atoms or a classical numerical simulation is the fact that the number of states of the gauge field, associated to links between lattice sites, must be finite. This problem has been tackled in different ways in literature. Sticking to $U(1)$ theories, it is possible to represent gauge fields by means of spin variables~\cite{simul2,rico2016}, considering the so called quantum link model~\cite{qlm1,qlm2,qlm3,qlm4wiese}. Otherwise, one can try to discretize the gauge group by replacing  $U(1)$ with $\mathbb Z$, which however admits only infinite-dimensional representations. In such a case, the Hilbert space describing the gauge degrees of freedom must be truncated~\cite{KCB, buyens}.
These approaches allows for the investigation of questions that are traditionally difficult to analyze, such as the emergence of quantum phase transitions, non-perturbative phenomena and dynamical aspects~\cite{simul2,simul6,BCJC,rico2016,montangero2015,schwinger_mps,buyens_prx,buyens,kuno,park}. 

We adopt here the scheme presented in~\cite{Erc}, where we considered a lattice version of the staggered-fermion Schwinger model~\cite{Sch1962}, in which the $U(1)$ gauge degrees of freedom are given in terms of the possible finite dimensional representations of the corresponding  Weyl group~\cite{NEFMPP}, thus yielding a discrete and finite implementation through the group $\mathbb{Z}_n$. In this approach, a systematic control of finite-$n$ effects is possible and a rigorous approximation of lattice $U(1)$ quantum electrodynamics in the large-$n$ limit becomes available to investigation. A rigourous investigation of this limit can be found in~\cite{Erc}, where the static phase diagram of this model has been studied,   while in~\cite{magn1,magn2} a variant of it has been shown to admit topological phases.  
In this article, our goal is to study out-of-equilibrium properties of discrete $\mathbb{Z}_n$ gauge models, which are relevant for comparison with experimental realizations of quantum simulators, that can be implemented only with a finite number of sites and, typically, a finite number of gauge degrees of freedom.  We will focus on the analysis of two relevant phenomena:
\begin{itemize}
\item[i)] the production of virtual particle/anti-particle pairs out of the Dirac sea vacuum, induced by quantum fluctuations; 
\item[ii)] dynamical effects of confinement, such as the string breaking mechanism. 
\end{itemize}

We will see that both these phenomena depend on the values of the two parameters that enter the Hamiltonian, namely the fermionic mass $m$ and the gauge coupling $g$, that will be defined and discussed in detail in Sect.~1. In particular, by means of simulations based on the Density Matrix Renormalization Group (DMRG) algorithm, we will examine the evolution of the system for a wide range of the parameters, showing that the dynamical behaviour of the model strongly deviates from the usual relaxation properties which are expected to be found in a many-body non-integrable system, resulting in stable and/or recurrent evolution of interesting physical quantities. This shows that confinement and a slow dynamics are not specific features of the $U(1)$ Schwinger model, but of the whole class of discrete lattice models we consider in this paper, which might be relevant for the description of future experiments with Rydberg atoms~\cite{simone}.
We remark also that similar results have been obtained in different many-body systems, such as non-integrable spin-chain models~\cite{tak} or constrained Hamiltonians~\cite{scar2} . 

This article is organized as follows.
In Sect.~\ref{sec:model}, we review the discretized version of the Schwinger model for 1+1 dimensional QED, that is accommodated on a one-dimensional lattice and endowed with a $\mathbb Z_n$ symmetry. This model, which depends on two physical parameters, namely the fermionic mass $m$ and the gauge coupling $g$, exhibits a quantum phase transition at $m_c = -0.33$, belonging the Ising universality class. For $m>m_c$ the ground state of the model is in a confined phase, in which elementary excitations above the Dirac sea vacuum are of mesonic type. On the other hand, for $m<m_c$, a symmetry is broken and a nonvanishing average electric field appears in the ground state.
In Sect.~\ref{sec:pair}, we fix the gauge coupling $g$ and set up a quench protocol to simulate spontaneous pair production occurring in absence of an external electric field, by starting form the Dirac sea vacuum and quenching the mass $m$ from an infinite to a finite value. We analyze this phenomenon by looking at the dynamical evolution of several physical quantities of interest, such as particle density, entanglement entropy, and density correlation functions. We will see that, contrary to what is found in many other integrable and non-integrable models, the production of correlated particle/antiparticle pairs is strongly suppressed when we consider system parameters deep in the confined phase.
In Sect.~\ref{sec:pairfield}, we examine the phenomenon of pair production induced by the presence of an external electric field, by working in the regime of strong confinement (large $m$). We find an agreement between our simulations and old predictions by Schwinger~\cite{Sch51} for the rate of pair production. Also, by examining the time evolution of entanglement, we conclude that the formation of mesonic excitations is stimulated by a dynamical effect due to the presence of the external field, of a different nature from the one emerging in the spontaneous case, examined in the Sect.~\ref{sec:pair}.
Finally, in Sect.~\ref{sec:string}, we will consider the real time evolution of a string excitation, spanning a wide range of both parameters $(m,g)$. We observe that the string breaks into mesons, thus giving rise to the so-called string-breaking mechanism, only in a weak confinement regime. Vice versa, deep in the confined regime, the strings remain localized and are apparently stable.

\section{The model}
\label{sec:model}

\begin{figure}
\centering
\includegraphics[width=0.4\textwidth]{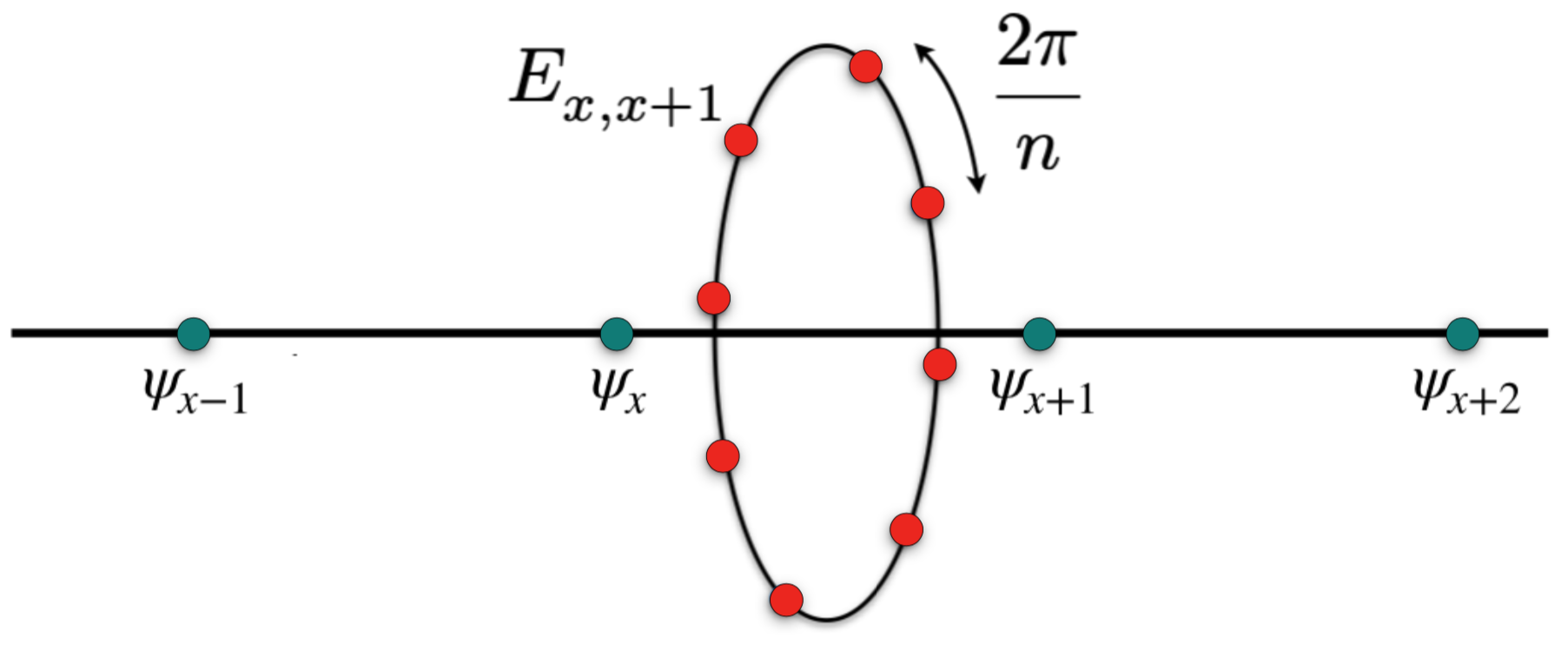}
\caption{$\mathbb Z_n$-discretization of QED in 1+1 dimensions, with fermionic matter $\psi_{x}$ living on sites $x \in \mathbb{Z}$ and electric field $E^{(n)}_{x,x+1}$ living on links between adjacent sites, described by a periodic discrete variable with a step $2\pi/n$. }
\label{fig:zn} 
\end{figure}

We start from the Hamiltonian of quantum electrodynamics (QED) in one spatial dimension:
\begin{equation}
H = \int \d x \bigg\{ \psi^{\dagger}\gamma^0 \left[ -\gamma^1 (\ii\partial_1 + g A) + m \right] \psi + \frac{E^2}{2} \!\bigg\} ,
\label{eq:Schwinger}
\end{equation}

\begin{equation}
\label{EAcont}
[E(t,x),A(t,x')]=\ii\delta(x-x').
\end{equation}
The Hamiltonian (\ref{eq:Schwinger}) must be complemented with the enforcement of Gauss's law
\begin{equation}
G(x) \equiv \partial_1 E(x) - g \psi^{\dagger}(x) \psi(x)  \approx 0 
\label{gauss}
\end{equation}
in a weak sense, meaning that the dynamics is restricted to the subspace, called the physical subspace, of states $\ket{\Psi}$ for which $G(x)\ket{\Psi}=0$. 

As described in Ref.~\cite{NEFMPP}, to which we refer for further details, it is possible to discretize this model following two steps: 
\begin{itemize}
\item[i)] first, we perform a spatial discretization, in which the space continuum is replaced by a linear lattice of points with spacing $a$; in order to avoid the fermion doubling problem we adopt \cite{kogut1975hamiltonian,susskind1977lattice,rothe1992lattice,montvay1997quantum} the staggered fermion approach~\cite{kogut1975hamiltonian,susskind1977lattice} applied to the Schwinger model~\cite{rothe1992lattice,montvay1997quantum}; 
\item[ii)] second, by following the Schwinger-Weyl quantization scheme~\cite{weyl,SE},  we approximate the  gauge group $U(1)$ with the finite group $\mathbb{Z}_n$; this step is essential in order to work with a finite number of local degrees of freedom also for the gauge variables. 
\end{itemize}

\begin{figure}
\centering
\includegraphics[width=0.48\textwidth]{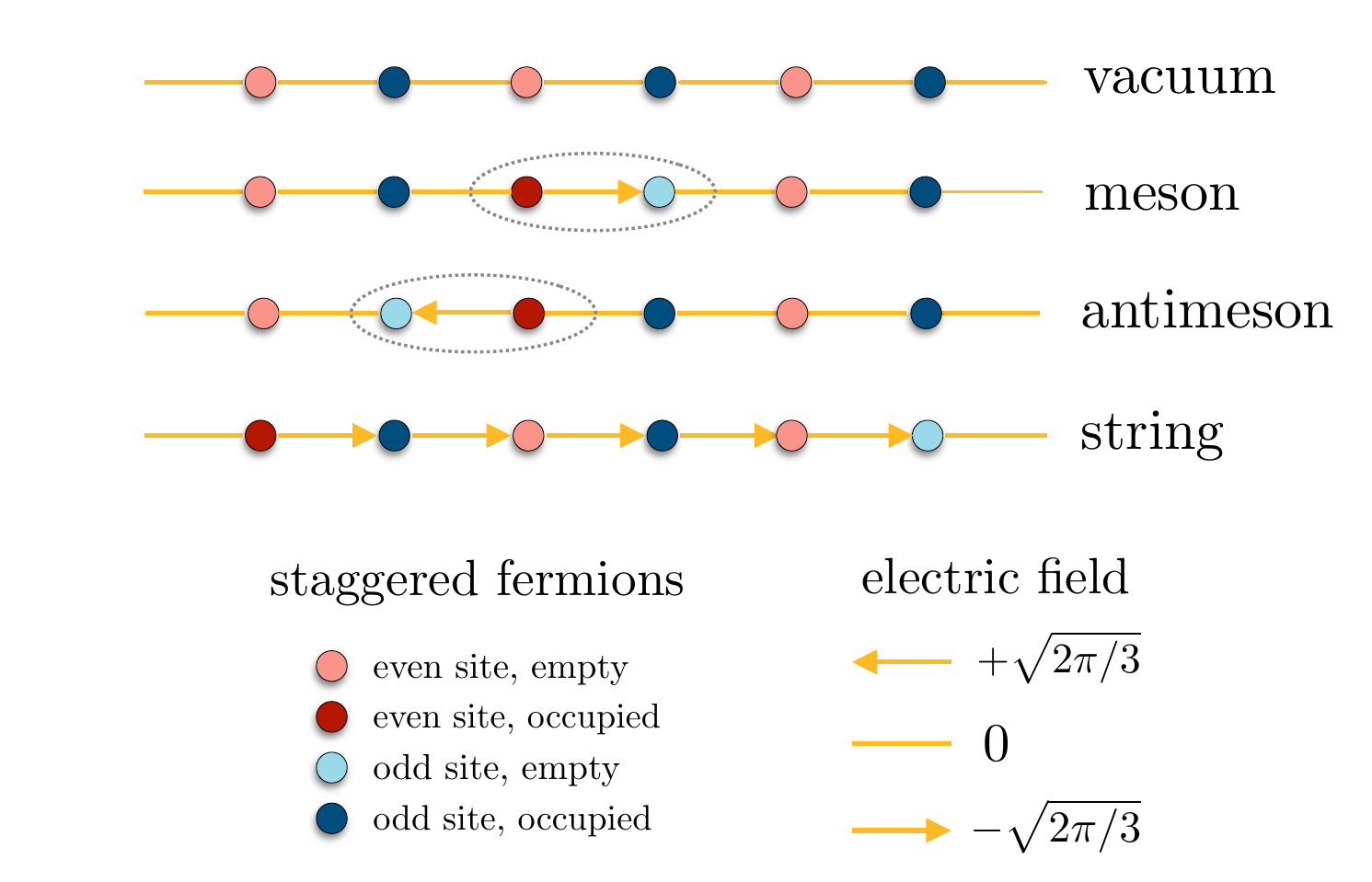}
\caption{Pictorial representation of notable configurations in a $Z_3$ gauge theory. All the represented cases are consistent with Gauss's law.}
\label{fig:2} 
\end{figure}

The discretized Hamiltonian reads~\cite{NEFMPP}:
\begin{eqnarray}
H & = & -\sum_{x}(\psi_{x+1}^{\dagger}U_{x,x+1}\psi_{x}+h.c.)\nonumber \\
 & & + \; m\sum_{x}(-1)^{x}\psi_{x}^{\dagger}\psi_{x}+\frac{g^{2}}{2}\sum_{x}E_{x,x+1}^{2},
 \label{HamQED2}
\end{eqnarray}
with $x$ ($1\leq x\leq N$) labelling the sites of a one-dimensional lattice of spacing $a=1$. Here, the one-component spinor is represented by the creation and annihilation operators $\psi_{x}^\dagger$ and $\psi_{x}$, defined on each site $x$ and characterized by a staggered mass $(-1)^x m$, while 
the gauge fields are defined on the lattice links $(x,x+1)$ through the pair of operators $E_{x,x+1}$ (electric field) and $U_{x,x+1}$ (unitary comparator).
The gauge field operators act on the $n$-dimensional Hilbert spaces ${\cal H}_{x,x+1}$, attached to each link and spanned by the orthonormal bases $\{ |v_k \rangle_{x,x+1} \}_{0\leq k\leq n-1}$, that diagonalize the local electric field:  
\begin{equation}
\label{eigenvn}
E_{x,x+1} = \sum_{k=0}^{n-1} \sqrt{\frac{2\pi}{n}}\bigg(k-\frac{n-1}{2}+\phi\bigg) |v_k \rangle_{x,x+1} \langle v_k | .
\end{equation}
Here, a non-zero value of the angle $\phi$ entails the presence of a constant background field, which in turn corresponds to charges at the boundary of the chain. The unitary comparator, instead, acts as a cyclic ladder operator: 
\begin{align}
\label{cyclic}
U_{x,x+1} |v_k\rangle_{x,x+1} & = |v_{k+1}\rangle_{x,x+1} \quad \mbox{for } k<n-1, \\
U_{x,x+1} |v_{n-1}\rangle_{x,x+1} & = |v_{0}\rangle_{x,x+1} .
\end{align}
Gauss's law is implemented by requiring that the physical states belong to the null space of the operators
\begin{equation}
\label{Gx}
 G_x \equiv \psi_x^{\dagger}\psi_x + \frac{1}{2}[(-1)^x-1] -(E_{x,x+1}-E_{x-1,x}),
\end{equation}
for all $x$.

Let us remark that, in one spatial dimension, the fermionic density $\psi_x^{\dagger}\psi_x$ and the local electric field completely determine each other up to a constant, which corresponds to the value of the electric field at one boundary. Thus, one can integrate out the gauge field in order to obtain an effective Hamiltonian in which only the matter fields appear, corresponding to a spin $1/2$ model with long-range interactions~\cite{surace}. On the other hand, one can eliminate the fermionic field to get an effective Hamiltonian that contains only the gauge variables. Within this approach, one thus obtains a local Hamiltonian with a $\mathbb Z_n$-symmetry, acting as (\ref{HamQED2}) when restricted to the physical subspace. The study of such Hamiltonian will be presented in a future work. Here, it is worth noticing that these ideas can be generalized also to higher dimensions~\cite{dalmon,cirac}. 

In order to clarify the relationship between the discretized Hamiltonian (\ref{HamQED2}) and the QED Schwinger model in the continuum, we briefly summarize their scaling properties, referring to~\cite{Erc} for all the details. 
The effects of lattice discretization can be studied by analyzing the scaling properties of the Hamiltonian with respect to a parameter $t$ that appears in front of the first term, the gauge invariant hopping. From the corresponding lattice Hamiltonian
\begin{align}
\label{HamQEDt}
h_t = & - \frac{t}{g^2a^2} \sum_x \!\left(\psi^{\dagger}_x
U_{x,x+1}\psi_{x+1}+\mathrm{H.c.} \right) \nonumber \\
& + \frac{2m}{g^2 a} \sum_x(-1)^x\psi_x^{\dagger}\psi_x+ \sum_x
E_{x,x+1}^2 \Biggr] ,
\end{align}
one can recognize that the scale of mass is fixed by $g^2a/2$, while the scale of the hopping parameter $t$ is set by $g^2a^2$; the standard lattice Schwinger model is recovered for $t=1$. From the above expression it is also evident that the parameter $t$ defines a rescaling of the lattice spacing according to: $a\rightarrow a/\sqrt{t}$. The semiclassical analysis and the numerical simulations presented in~\cite{Erc} show that the critical value of the mass scales as $m_c(t) = \alpha \sqrt{t}$ so that its value in the continuum can be recovered by finding: $m_c^{(cont)}= m_c(t=1) = \alpha$. 
Also, the discretization of the gauge group introduces a factor of $\sqrt{2\pi/n}$ in the definition of the electric field, see Eq.~(\ref{eigenvn}), which can be absorbed in the definition of the gauge coupling constant by setting $g \rightarrow g_n = g \sqrt{2\pi/n}$. Accordingly, we can actually write $m_c(t) = \alpha_n \sqrt{t}$ and recover the $U(1)$ limit by calculating the coefficient: $\alpha = \lim_{n\rightarrow\infty} \alpha_n \sqrt{n/2\pi}$. Let us remark that, according to the above discussion, numerical simulations which implicitly assume $a=1$, such as the ones presented in Sects.~\ref{sec:pair}--\ref{sec:pairfield},  are performed by setting $g= \sqrt{n/\pi}$.

In Ref.~\cite{Erc}, we analyzed the phase diagram of the discretized Hamiltonian (\ref{HamQED2}) showing that, for any $n$, it exhibits a phase transition at a critical value of the mass $m_c$ between a confined phase for  $m>m_c$, in which the ground state is given by a dressed Dirac sea vacuum, and a deconfined phase for $m<m_c$, in which the ground state is characterized by a nonvanishing average electric field, which entails a symmetry breaking. 

In our representation, the Dirac sea vacuum is obtained by filling up all odd sites (negative mass fermions) and leaving the even ones (positive mass fermions) empty, as shown in the first line of Fig.~\ref{fig:2}. In this case, Gauss's law (\ref{Gx}) is satisfied if the electric field is zero on any link, a fact that we represent in the figure with a non-oriented link. A meson (antimeson) is obtained by acting on the Dirac sea by moving one fermion on an odd site to the right (to the left), as shown in the second (third) line of Fig.~\ref{fig:2}. In this case, Gauss's law requires that the electric field on the connecting link is equal to $+\sqrt{2\pi/n}$ ($-\sqrt{2\pi/n}$), represented in the picture by an oriented link pointing to the right (left). In the fourth line of Fig.~\ref{fig:2} we also show another gauge invariant configuration, representing a string, in which the particle/antiparticle excitations that constitute a meson are moved farther away from each other, with the electric field being constant and nonvanishing on all the intermediate links. The real-time dynamics of such string configurations will be studied in the second part of this article. 

The phase transition belongs to the Ising universality class for all $n$, with the confined/deconfined cases corresponding respectively to the paramagnetic/ferromagnetic phases of the Ising model in a transverse field. Here, the quantum phase transition is driven by the value of the fermion mass (playing the role of the external transverse magnetic field) and the order parameter (the magnetization in the Ising case) is represented by the mean value of the electric field operator
\begin{equation}
\varSigma=\frac{1}{N}\sum_{x}\left\langle E_{x,x+1}\right\rangle ,
\end{equation}
or, equivalently, by the mean fermion density
\begin{equation}
\rho =\frac{1}{N}\sum_{x} \langle\frac{1}{2}\left[1-(-1)^{x}\right]+(-1)^{x}\psi_{x}^{\dagger}\psi_{x}\rangle ,
\end{equation}
which, in the thermodynamic limit, are nonvanishing only in the deconfined phase. The value of $m_c$ depends on the dimension $n$ used to discretized the $U(1)$ gauge group. In particular $m_c$ is positive/negative when $n$ is even/odd. As discussed in~\cite{Erc}, the negative value of the critical mass for even $n$ is related to the fact that the electric field cannot vanish on a link. For large odd $n$, $m_c$ approaches the value $ -0.33$ of the continuum theory. In addition, as expected in the Ising model, close to the transition point the first excitation has  conformal dimension $d=2$, corresponding to a kink-like (or domain wall) solution. 

Let us remark that while the Ising model in a transverse field is integrable for any value of the magnetic field, our model is more complicated and never integrable (except for the trivial case $g=0$), because of the gauge coupling between fermionic matter and electric field. In the Ising case, integrability is lost and effective interactions between domain-wall excitations are present only if one adds the coupling with an external uniform longitudinal field~\cite{Wu}. This effect can be mimicked in our model by introducing a background constant electric field.

The effects of such gauge-mediated interactions might be quite strong and will be studied by looking at real-time dynamical properties of our model in the next sections. The analysis will be performed by numerically studying the Hamiltonian (\ref{HamQED2}) with a $t$-DMRG algorithm, whose dynamical evolution is implemented through the Runge-Kutta method. Details about the precision of our algorithm are given in the Appendix. 

\section{Spontaneous pair production}
\label{sec:pair}

In this section, we will examine the phenomenon of spontaneous pair production in the $\mathbb{Z}_{n}$-Schwinger model for  $(1+1)$-dimensional QED, by simulating with our model the real-time dynamics of the Dirac sea vacuum, in absence of an external electric field. This effect has also been considered in other approaches~\cite{montangero2015,schwinger_mps, buyens_prx, buyens} and experimentally analyzed in a small system (4 qubits) of trapped ions~\cite{martinez2016}. 

\begin{figure}
\centering
\includegraphics[width=0.4\textwidth]{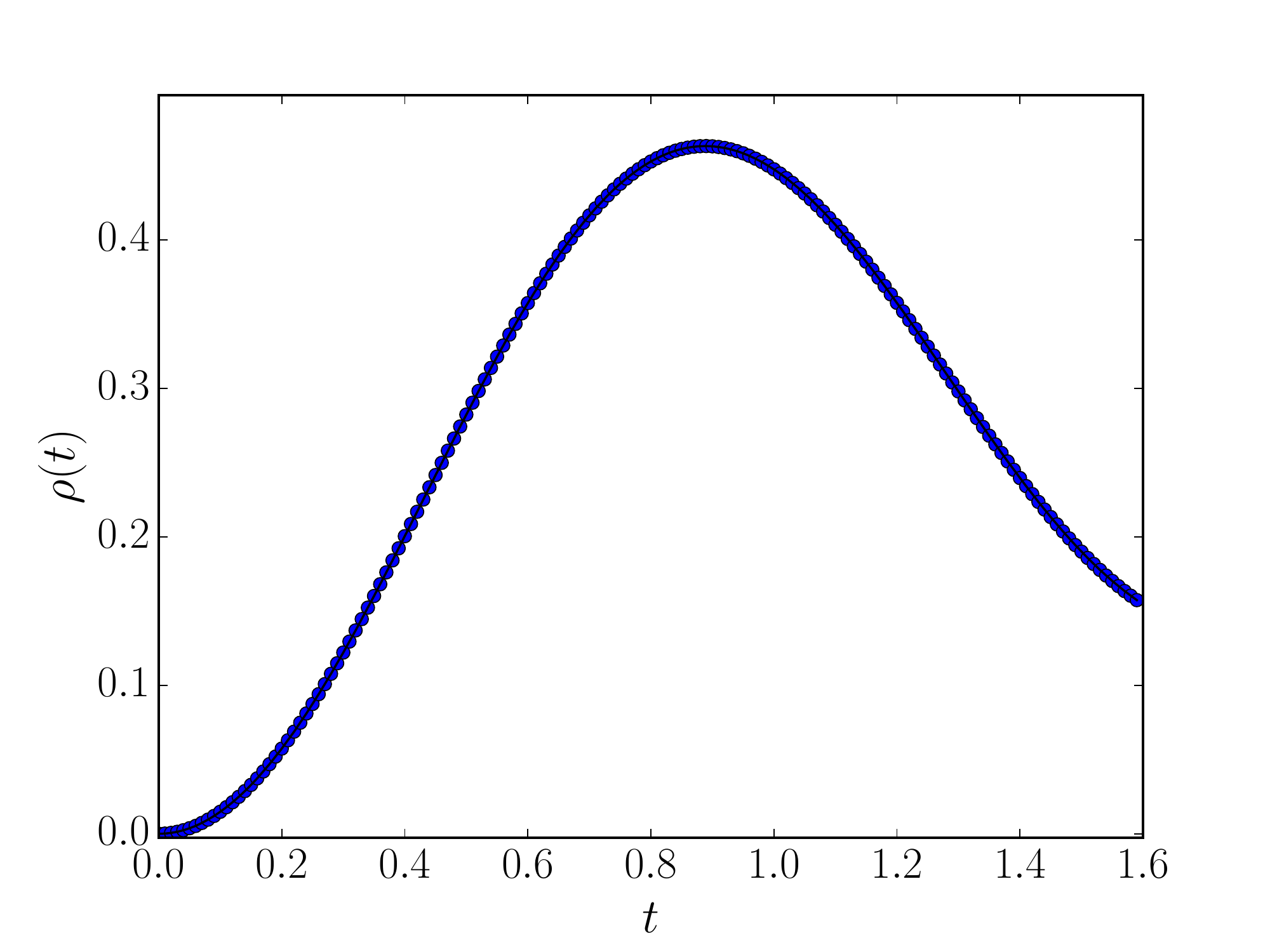}
\caption{Time evolution of the particle density for a system of $N=4$ sites, prepared at $t=0$ in the Dirac sea vacuum and evolving under the $\mathbb{Z}_3$ model Hamiltonian. Here and in all the following figures, time is  measured in units $[\text{energy}]^{-1}$. }
\label{fig:n4} 
\end{figure}

In order to test the stability of the Dirac vacuum with respect to spontaneous pair production, we prepare the system in the $m\to\infty$ ground state
\begin{equation}
\ket{\Psi_0(t=0)} = \ket{0}_{\mathrm{Dirac}} ,
\end{equation}
corresponding to the Dirac sea vacuum, and study its evolution under the action of the Hamiltonian with different values of the coupling constants $m,g$, with either $m>m_c$ or $m<m_c$. For completeness, we will perform our analysis for all values of $m$, both positive and negative, but we stress that the Dirac vacuum we take as initial state is highly excited for $m<m_c$, while it is very close to the true ground state in the confined phase and in particular when $m$ is large and positive. 
In the language of spin systems, this would be analogous to preparing an Ising system in the paramagnetic ground state, by setting the external transverse field $h=0$, and then suddenly quenching the Hamiltonian to a different value of $h$, with $h<1$ (ferromagnetic phase) or with $h>1$ (quenching into the paramagnetic phase)~\cite{tak}. 

Let us consider the $\mathbb Z_3$ case. To test the evolution of the Dirac sea vacuum after the quench, we shall numerically calculate three different quantities: mean particle density, entanglement entropy, and correlation functions.

\subsection{Mean particle density}

The first quantity we consider is the mean particle density  
\begin{equation}\label{rhodens}
\rho(t) =\frac{1}{N}\sum_{x} \langle \Psi_0(t) | \frac{1-(-1)^{x}}{2}+(-1)^{x}\psi_{x}^{\dagger}\psi_{x}| \Psi_0(t) \rangle , 
\end{equation}
evaluated on the evolved Dirac sea vacuum 
\begin{equation}
|\Psi_0(t) \rangle = \e^{-\ii t H} |\Psi_0(t=0) \rangle,
\end{equation}
where, since $\hbar=1$, time is measured in units of $[\text{energy}]^{-1}$.
The quantity~\eqref{rhodens} vanishes on the Dirac sea vacuum and takes the value $1$ on the state with the maximal number of mesons. 

To perform a first test on our model, we calculate $\rho(t)$ for a chain of $N=4$ sites, after a quench to $m=0.5$ and $g=\sqrt{6/2\pi}$, which correspond to the parameters used  in the experimental set up of Ref.~\cite{martinez2016}. Figure~\ref{fig:n4} shows our result: the density starts form zero and arrives very close to the value $1/2$, corresponding to the formation of one meson. After reaching a maximum, recombination effects bring the value down again, in perfect agreement with observations~\cite{martinez2016}.

\begin{figure}
\centering 
\subfigure[\label{fig:densita_m_grandi}]{\includegraphics[width=0.5\textwidth]{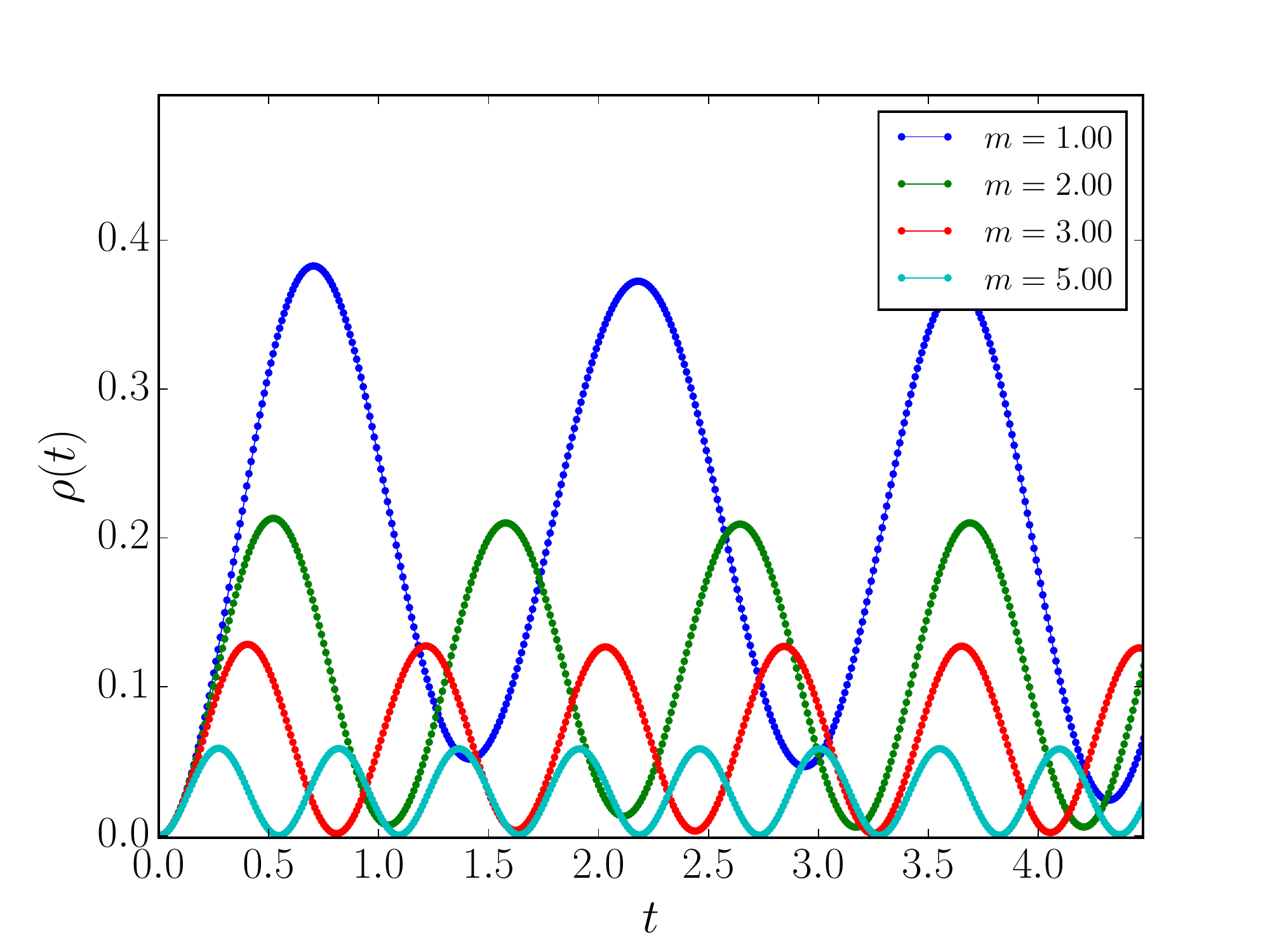}} 
\subfigure[\label{fig:densita_m_grandi_negativi}]{\includegraphics[width=0.5\textwidth]{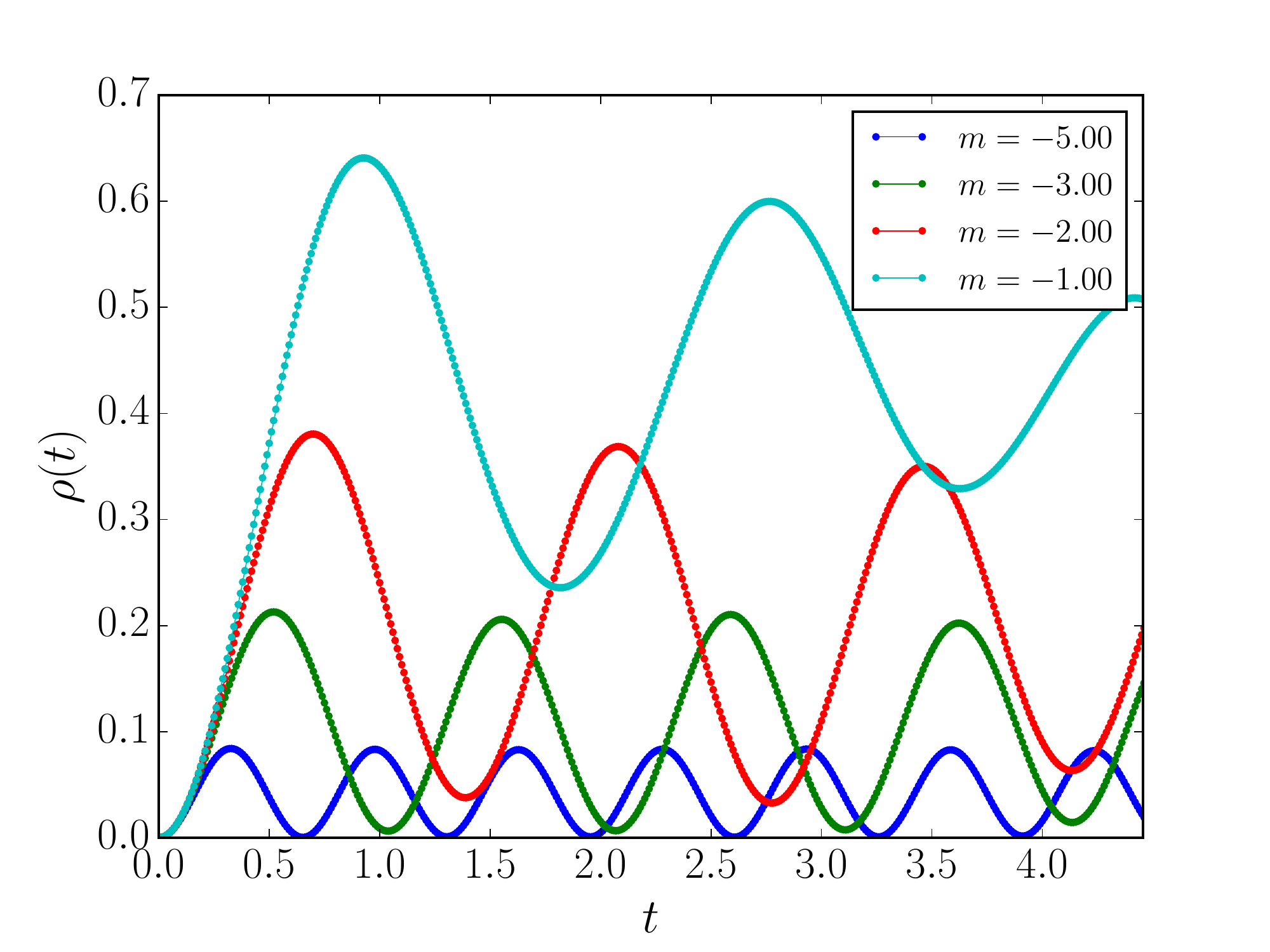}}
\subfigure[\label{fig:densita_m_vicini_a_zero}]{\includegraphics[width=0.5\textwidth]{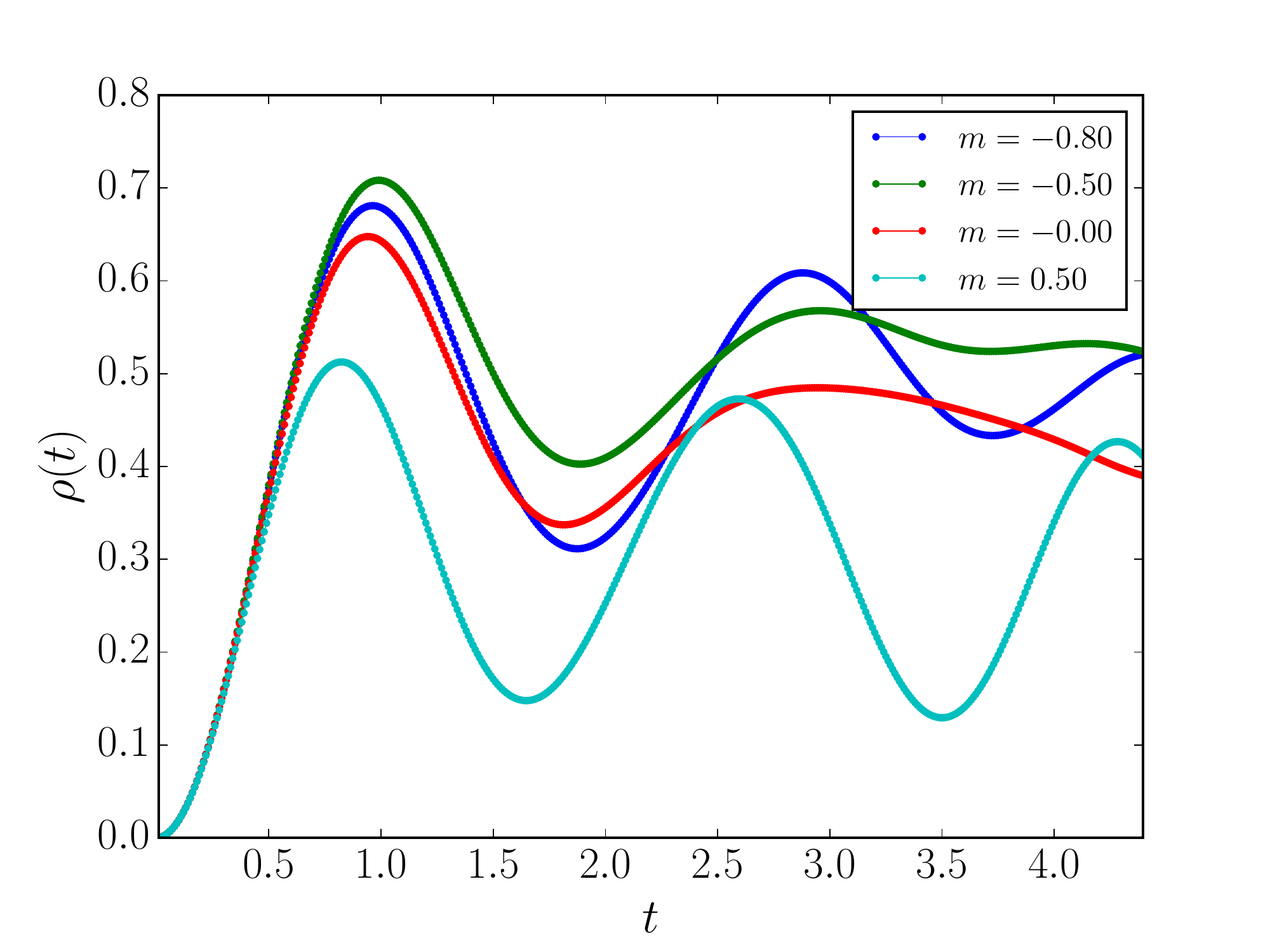}}
\caption{$\mathbb{Z}_3$-model. Time evolution of particle density in a system initially prepared in the Dirac sea vacuum and quenched to different values of mass, at fixed $g=\sqrt{3/\pi}$ and $N=40$: (a) $m\geq 1.0$,  (b) $m\leq -1.0$, (c) $-1.0<m<1.0$. } \label{fig:densita_m}
\end{figure}

More generally, we set $g=\sqrt{6/2\pi}$ and quench to different values of $m$. The temporal evolution of $\rho$ is shown in Fig.~\ref{fig:densita_m} for (a) large positive values of the mass ($m\geq 1.0$), (b) large negative values of the mass ($m\leq -1.0$), (c) small absolute values of the mass ($-1.0<m<1.0$). We clearly see that for very large values of the mass (both positive and negative), $\rho(t)$ periodically oscillates between zero and a rather small value, due to a small pair production rate and to strong recombination effects: thus, the state of the system periodically returns to the Dirac sea vacuum. On the contrary, for smaller values of $|m|$, the density increases rapidly up to values close to $1$ (corresponding to the mesonic ground state), to start then oscillating because of recombination effects, now around a large value; these oscillations do not bring the particle density back to zero, showing that the evolved state does not return to the Dirac sea vacuum. This is particularly evident for quenches to values of the mass close to the critical one, $m_c \simeq -0.33$.

\begin{figure}[t!]
\centering
\subfigure[\label{fig:picco}]{\includegraphics[width=0.43\textwidth]{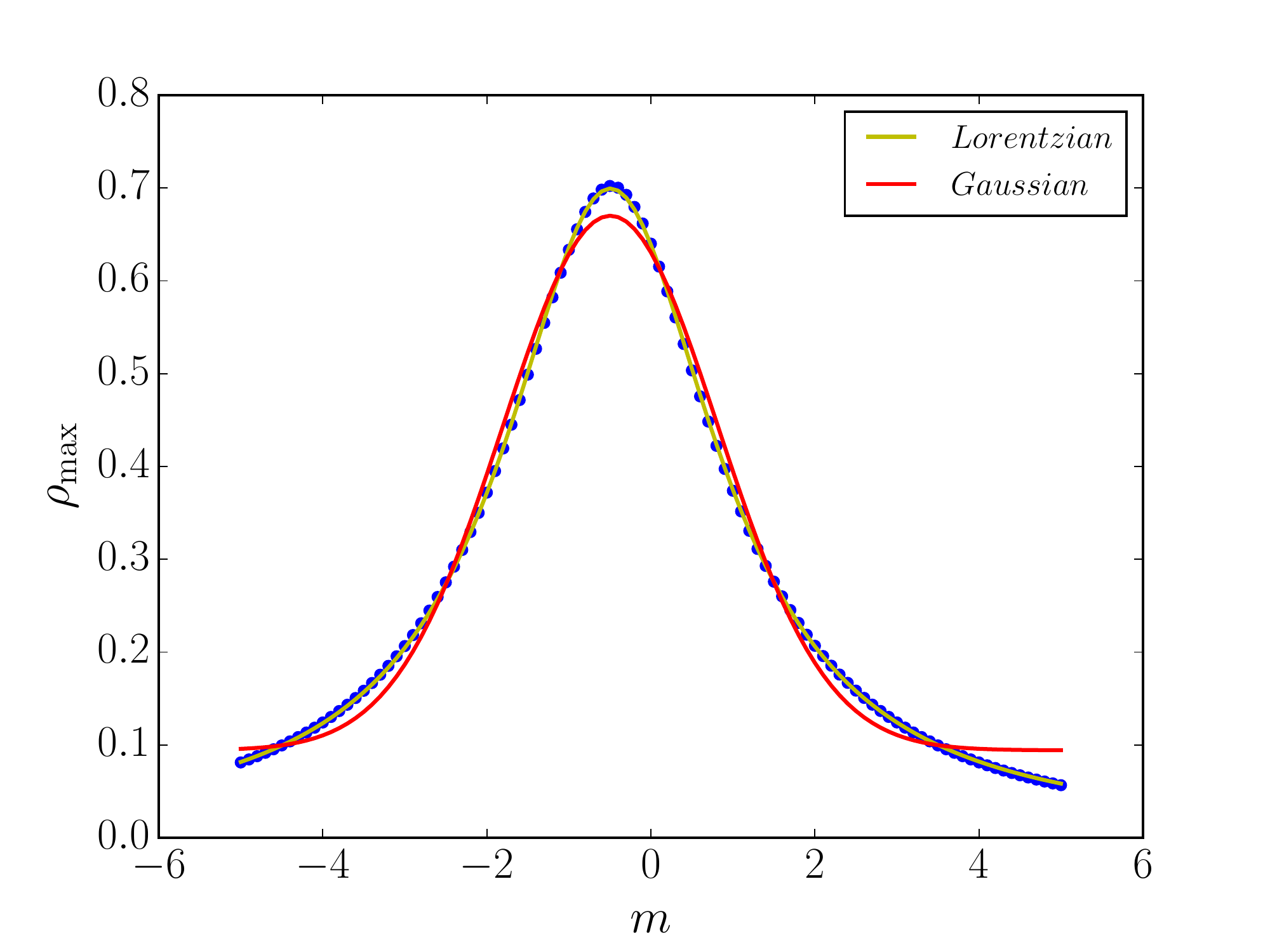}}
\subfigure[\label{fig:periodo}]{\includegraphics[width=0.43\textwidth]{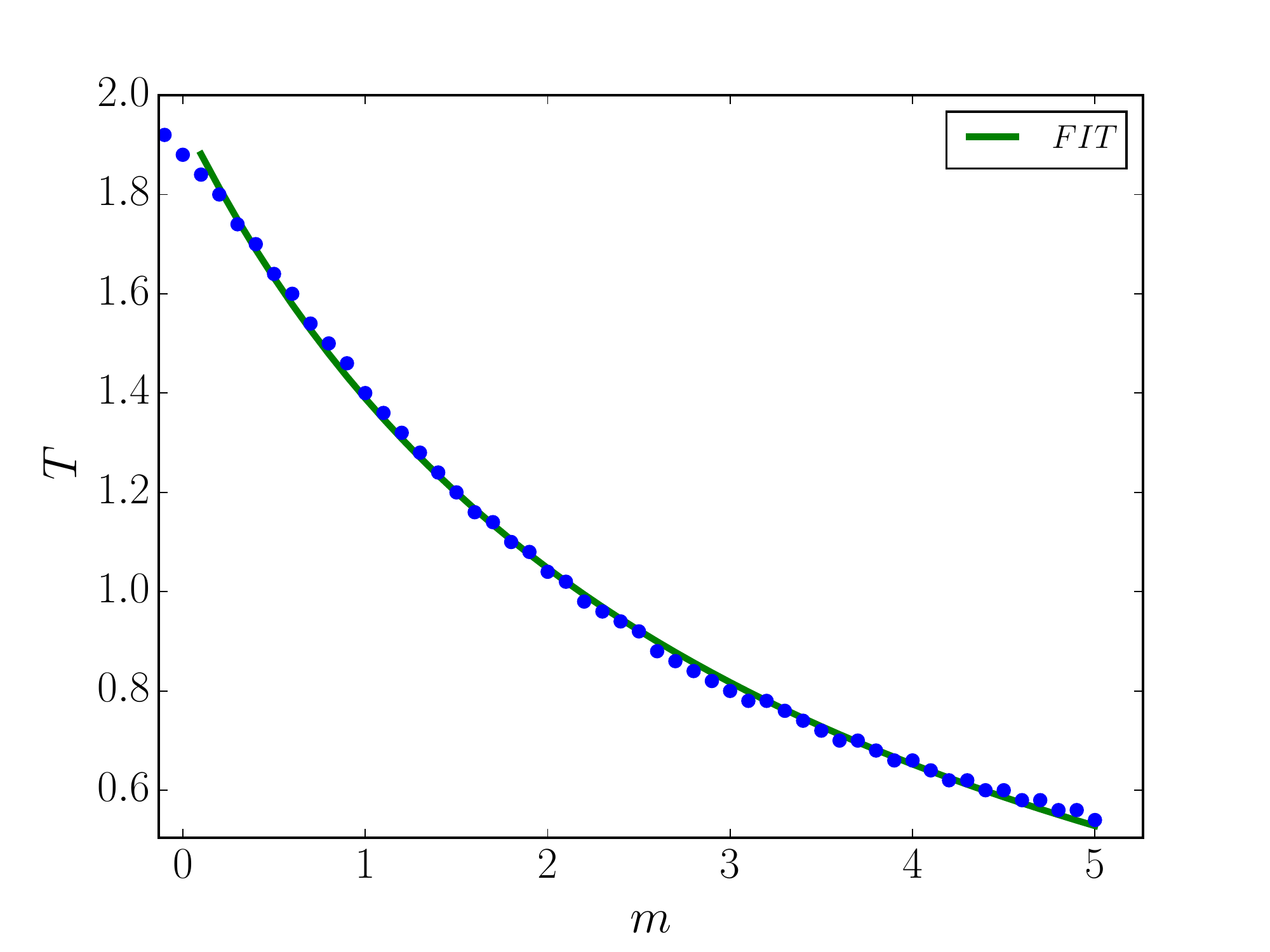}}
\caption{Analysis of the Dirac sea vacuum stability for $g=\sqrt{3/\pi}$ and varying $m$. (a) Value of the first maximum of $\rho(t)$. The fits are performed with a Lorentzian and a Gaussian distribution as described in the text. (b) Period of the oscillations after the first maximum, for $m>m_c$. 
The fit yields  $T = 1/(a m+b)$, with $a=0.26$ and $b=0.42$. 
} 
\label{fig:analisi}
\end{figure}

Since the time of our simulations is limited by the precision of the algorithm, as we explain in the Appendix, we are not able to extract the asymptotic plateau value of the density and the exact period of the oscillations. However, from the numerical data shown in Fig.~\ref{fig:densita_m} as a function of $m$, we can: i) extract the value of the first maximum of $\rho(t)$ as shown in Fig.~\ref{fig:picco}; ii) obtain an estimate of the period $T$  of the oscillations, based on the difference between the times of the second and first maxima, as shown in Fig.~\ref{fig:periodo}. It is clear that the pair population is peaked around the phase transition value. The red/yellow curves of Fig.~\ref{fig:picco} show a Lorentzian/Gaussian fit resepectively. There is a good agreement with the first one,  performed via the function $A [ \gamma^2 / (\gamma^2 + (m-m_0)^2] + c$, with $m_0= -0.493,\;  A=0.695,\;  \gamma=1.594,\; c= 0.00435$, with errors of the order of $10^{-3}-10^{-4}$. We should stress, however, that if we try a fit of the tails only, where the effect of confinement is strong, both a Lorentzian and a Gaussian curve work very well. As for the period $T$ of the oscillations, for $m>m_c$, we see that the data are very well described by the continuous line in Fig.~\ref{fig:periodo}, which corresponds to the best fit obtained with the functional form $T(m) = 1/(a m+b)$. This form is expected from a first-order approximation, according to which the period might be estimated from the inverse of the energy difference between the Dirac sea energy ($E_{\mathrm{Dirac}}/N= -m$) and the energy of the mesonic state ($E_{\mathrm{meson}}/N= m + g^2/2$), yielding  $a= 2, \; b=g^2/2\simeq 0.48$. The best fit confirms such a functional dependence, but with different values of the parameters ($a=0.26$ and  $b=0.42$), indicating strong quantum corrections.

\begin{figure}
\centering{}\includegraphics[scale=0.42]{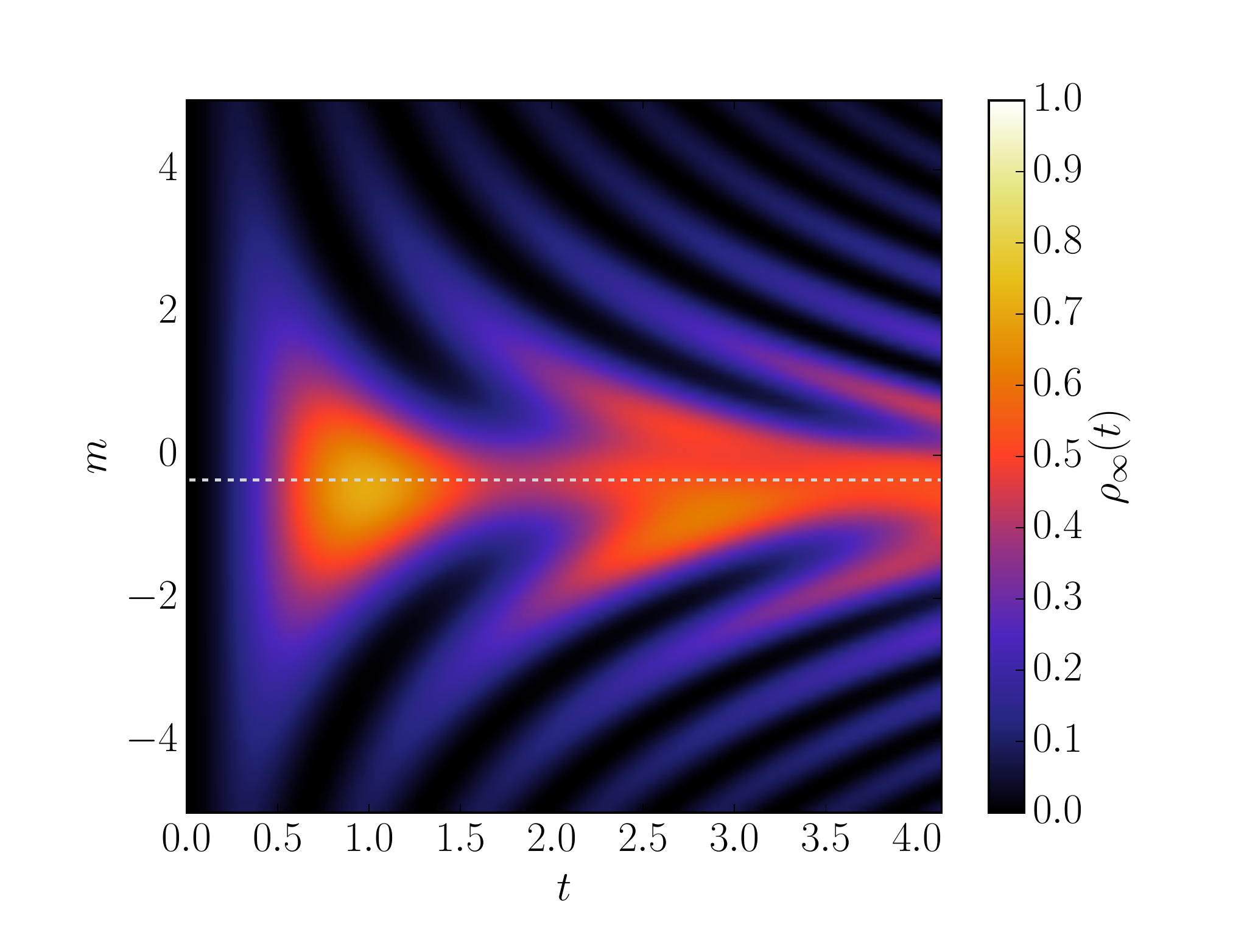}\caption{\label{fig:densita_estrapolata_ogni_m} Contour plot of the asymptotic particle density $\rho_\infty (t)$, obtained by extrapolating different finite-$N$ results, for $g=\sqrt{3/\pi}$ and quenched mass $m\in[-5,+5]$. The horizontal dashed line corresponds to $m=m_c$.}
\end{figure}

In the Appendix, we give details on the finite size scaling analysis we performed. To summarize our results, we show in Fig.~\ref{fig:densita_estrapolata_ogni_m} the contour plot of the density $\rho_\infty (t)$ extrapolated for $N\rightarrow \infty$, in the whole range of the quenched mass $m\in[-5,+5]$. We can conclude that the evolution brings the system towards high values of charge density when the mass is close to the critical value $m_c$, whereas pair production is strongly suppressed for large values of $|m|$.

\subsection{Entanglement entropy}

\begin{figure}
\centering
\subfigure[\label{fig:entropia_grandi_m}]{\includegraphics[scale=0.38]{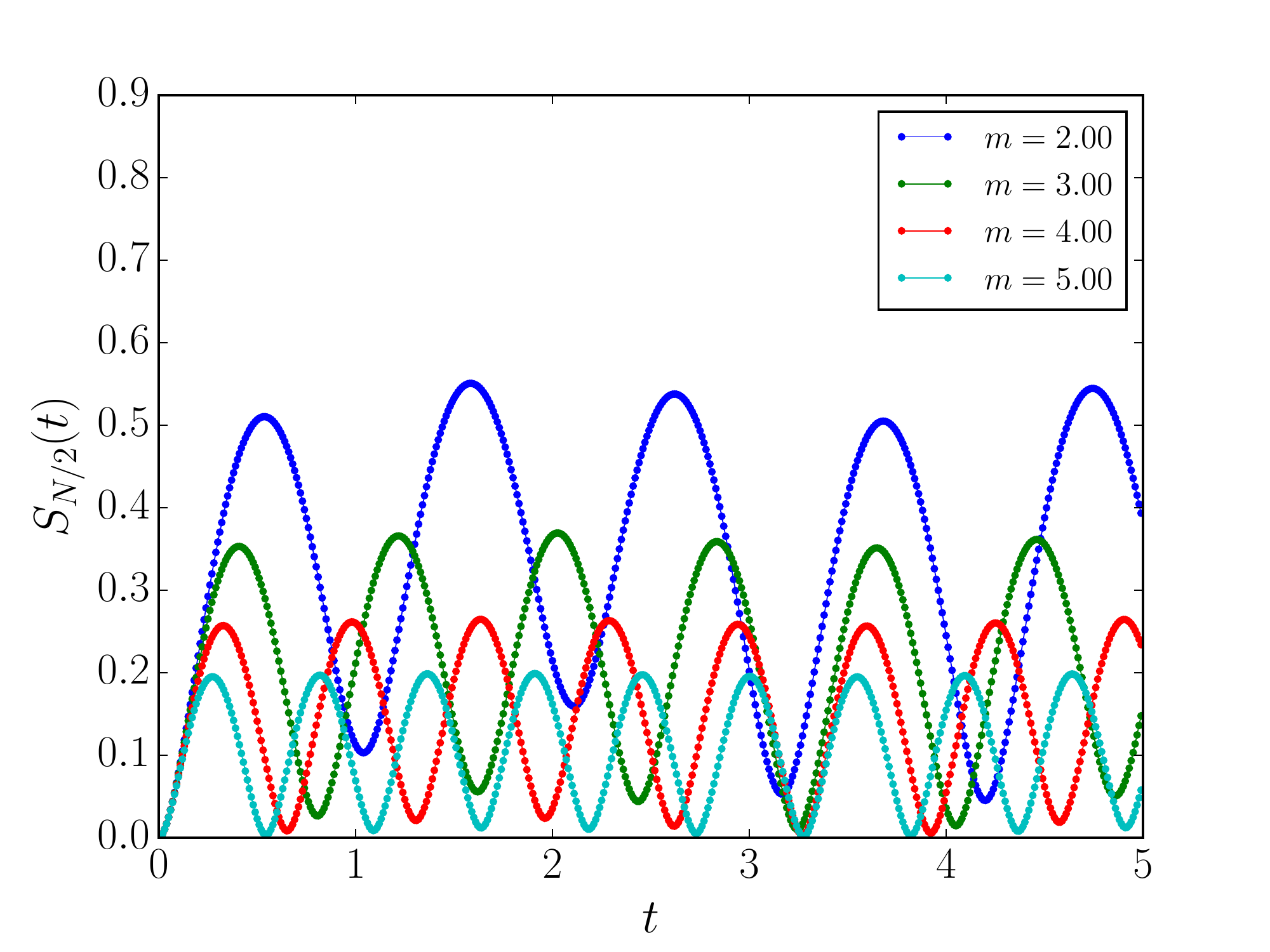}}
\subfigure[\label{fig:entropia_m_negativi}]{\includegraphics[scale=0.38]{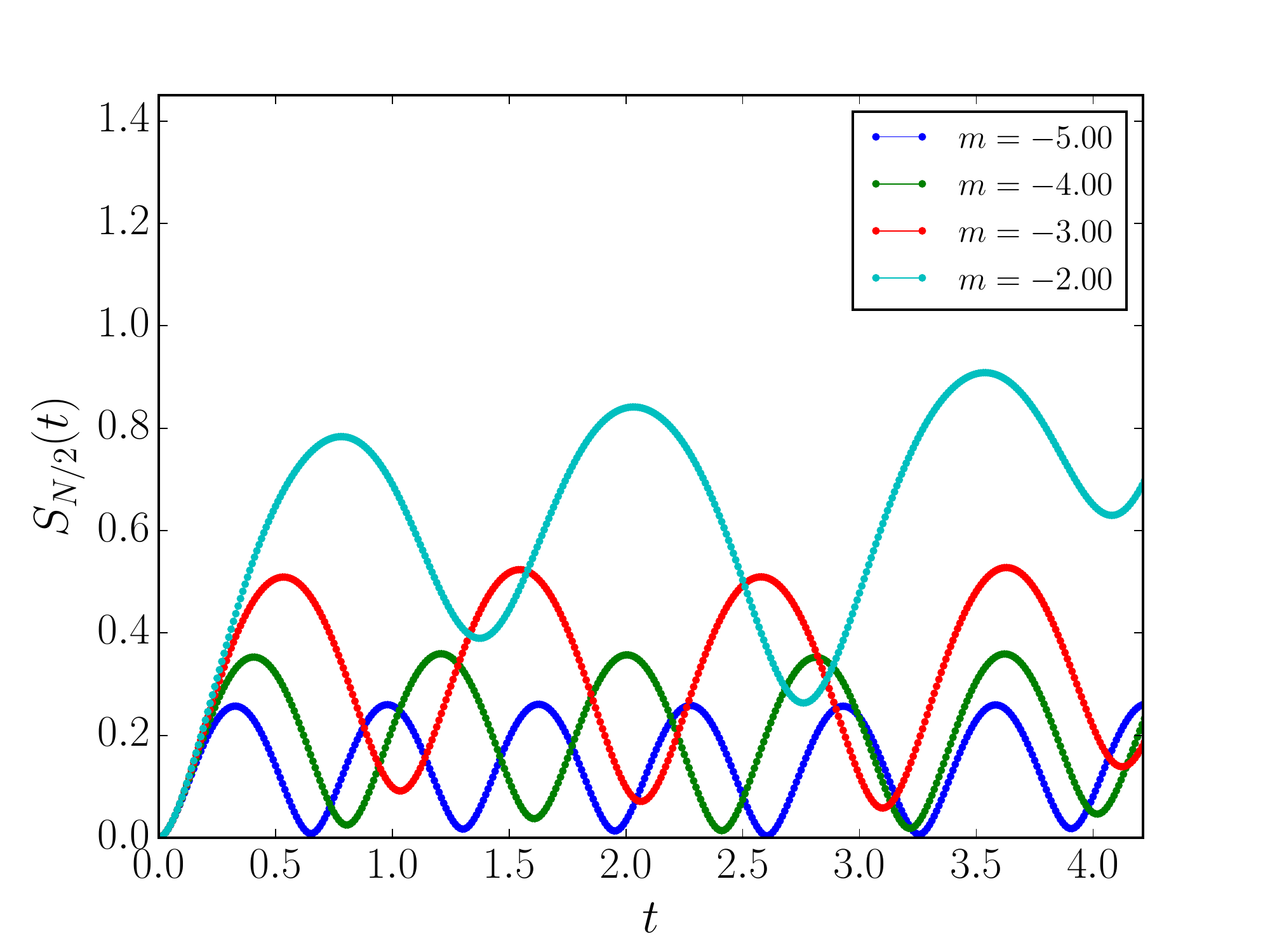}}
\caption{ $\mathbb{Z}_3$-model.  Time evolution of the half-chain entanglement $S_{N/2}(t)$ of a system initially prepared in the Dirac sea vacuum, for $g=\sqrt{3/\pi}$ and (a) $m\geq 2.0$, (b) $m\leq - 2.0$, (c) $-2.0<m<2.0$. }
\end{figure}

The second quantity we analyze is the time evolution of the half-chain entanglement entropy
\begin{equation}
S_{N/2}(t)=-\mathrm{Tr_{A}}\left\{ \rho_{A}(t)\log_{2}\left[\rho_{A}(t)\right]\right\} ,
\end{equation}
where the chain has been partitioned in two subsytems $A$ and $B$, consisting of the first and last $N/2$ sites of the chain, respectively. We recall that this quantity has attracted much attention in the literature, since it yields a good quantifier for bipartite entanglement in quantum many body systems, and displays universal properties in the vicinity of a phase transition~\cite{Special1, Special2}.

\begin{figure}
\centering
\subfigure[\label{fig:entropia_m_vicini_a_zero}]{\includegraphics[scale=0.38]{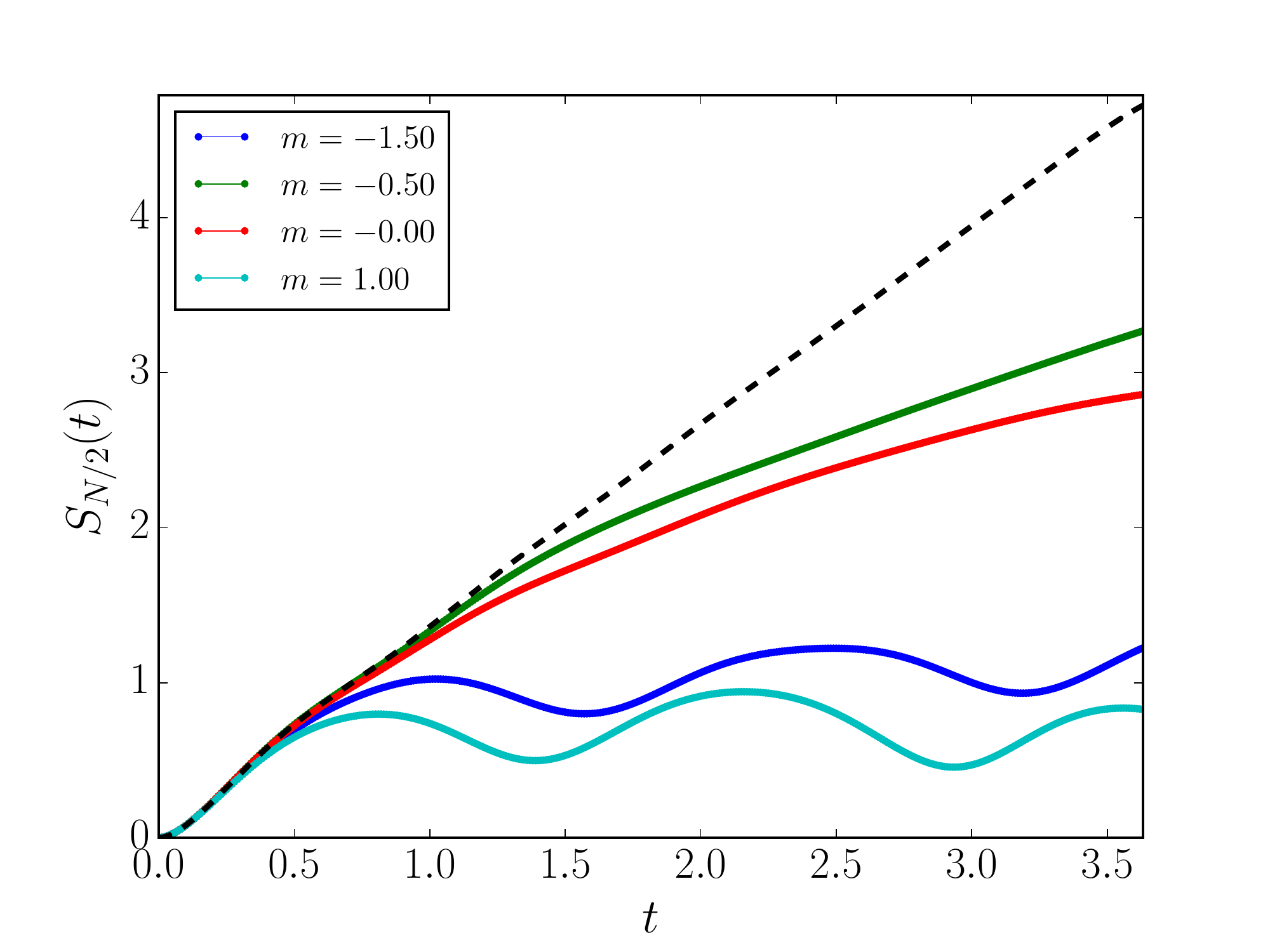}}
\subfigure[\label{fig:entropia_log}]{\includegraphics[scale=0.38]{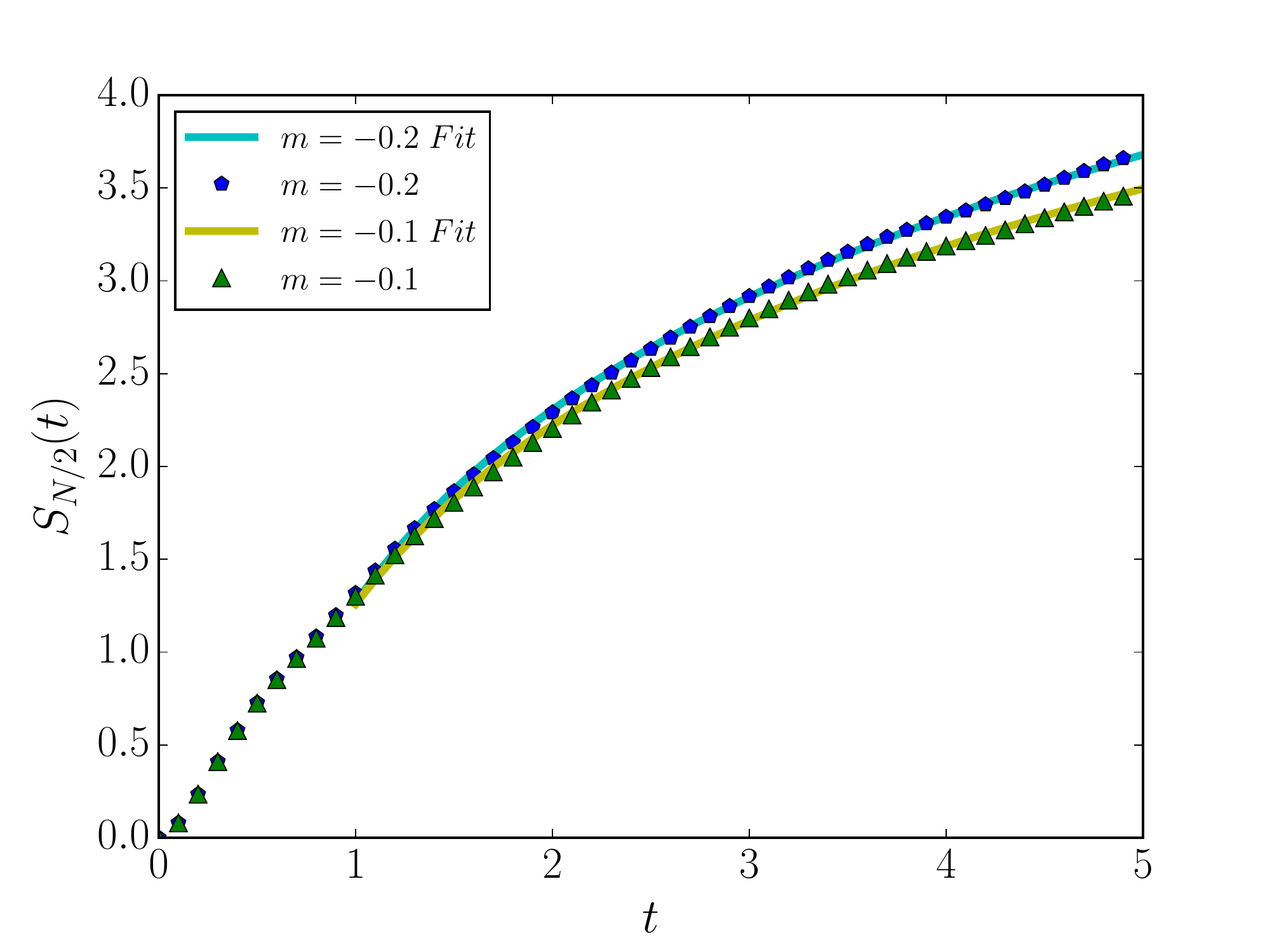}}
\caption{ $\mathbb{Z}_3$-model. (a) Time evolution of the half-chain entanglement $S_{N/2}(t)$ of a system initially prepared in the Dirac sea vacuum, for $g=\sqrt{3/\pi}$ in the range $-2.0<m<2.0$.  The black dashed line represents the free-fermion case $(m=0,g=0)$. (b) Same as in (a) but with $m=-0,1, -0,2$.  The  continuous lines are logarithmic fits}. 
\end{figure}

The initial Dirac sea vacuum is separable, hence $S_{N/2}(t=0) =0$.
Figures~\ref{fig:entropia_grandi_m}--\ref{fig:entropia_m_negativi} show some examples of $S_{N/2}(t)$ for large (positive and negative) values of $m$ and $g=\sqrt{3/\pi}$. The entanglement entropy displays an oscillatory behaviour, reaching a maximum at relatively small values of $m$ (e.g.\ $S_{N/2}^{\max} \sim 0.5$ for $m=2.0$ or $m=-3.0$). The period of these oscillations increases as $m$ decreases and, at the same time, the entropy reaches higher values, of the order of unity, for small $m$, as one can observe in Fig.~\ref{fig:entropia_m_vicini_a_zero}.  The qualitative explanation for the $m$-dependence of the period is the same as that given in Fig.~\ref{fig:analisi}.

For $m$ close to the critical value $m_c$ (see for example the data for $m=0.0, -0.5$), the entropy increases much faster and monotonically. Thus, we can conclude that when the phenomenon of pair production is dominant, quantum correlations between the two parts of the chain are larger due to the fact that the spontaneously created entangled particle/antiparticle pairs spread along the chain. 
This effect has been extensively studied in free fermionic models~\cite{key-1}, where entanglement can be shown to increase linearly with time, at least as long as one can assume a maximum speed of propagation (the Lieb-Robinson bound~\cite{LR}). 
If disorder and/or long-range interactions are not considered, this behavior has also been confirmed in several integrable and non-integrable models~\cite{int1,int2,int3,int4,int5,int6,int7}. In Ref.~\cite{tak}, the time evolution of the half-chain entanglement entropy has been studied for the Ising model with both a transverse and a longitudinal field, which admits a confined phase~\cite{Wu}, showing that the growth of the entanglement entropy is strongly reduced for quenches within the confined (ferromagnetic) phase. This corresponds to what we observe in our model~\cite{f1}. The black dashed line in Fig.~\ref{fig:entropia_m_vicini_a_zero} shows the linear behavior of the entanglement entropy in the free case. For small but non-zero values of $m$, we can clearly recognize a slow-down of the entanglement entropy growth, which is now well described by a logarithm: $S_{N/2}(t) \sim \log t$, as shown in Fig.~\ref{fig:entropia_log}. When we are deep in the confined phase, the system does not seem to evolve toward an equilibrium configuration, at least on the time scale of our simulations. Indeed, entropy is strongly suppressed and shows revivals, similar to what happens in models where the quantum scar phenomenon appears~\cite{scarexp,scar1,scar2}.

\subsection{Correlation functions}

\begin{figure}
\subfigure[\label{fig:g_05}]{\includegraphics[scale=0.4]{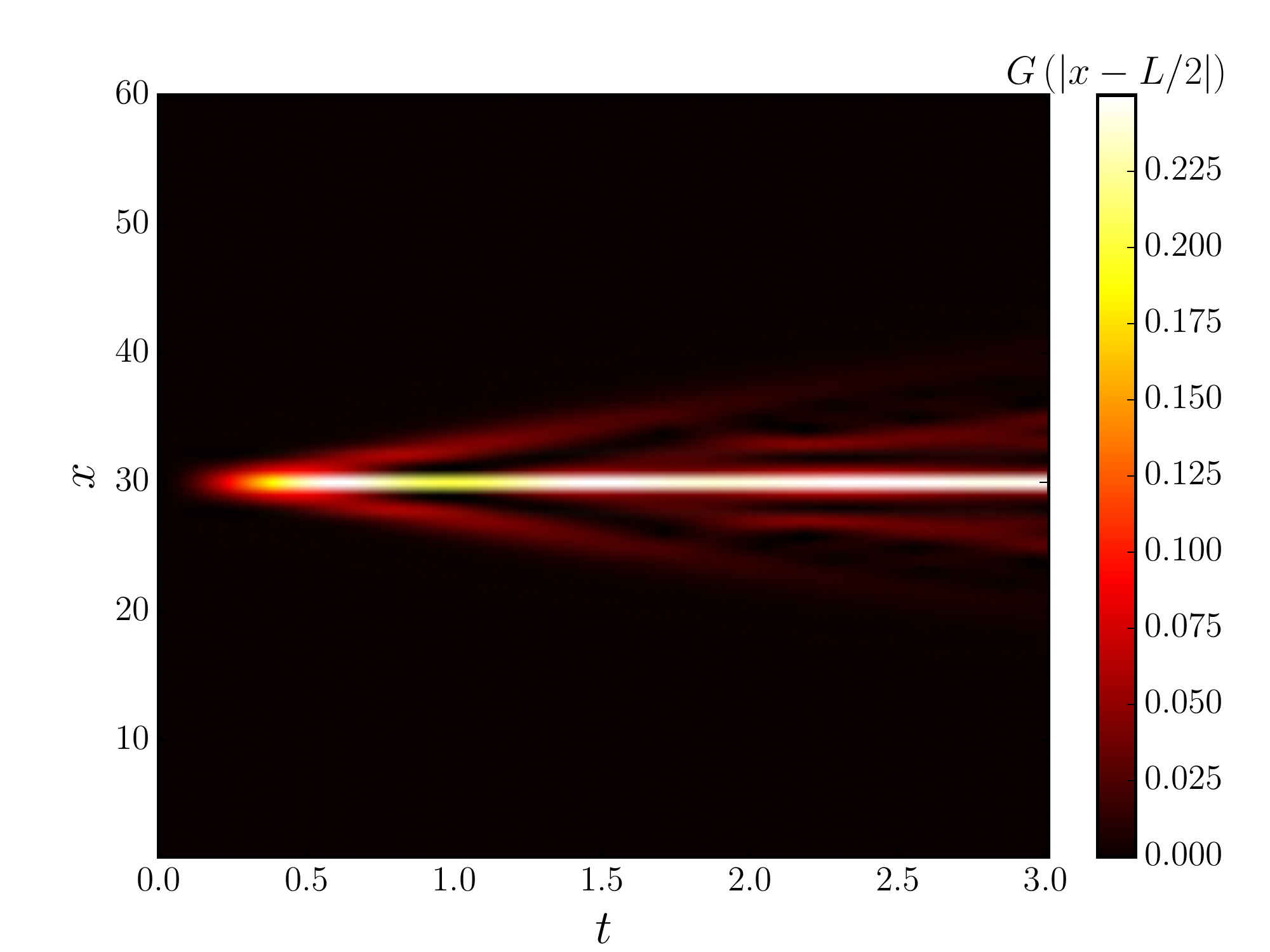}}
\subfigure[\label{fig:g_10}]{\includegraphics[scale=0.4]{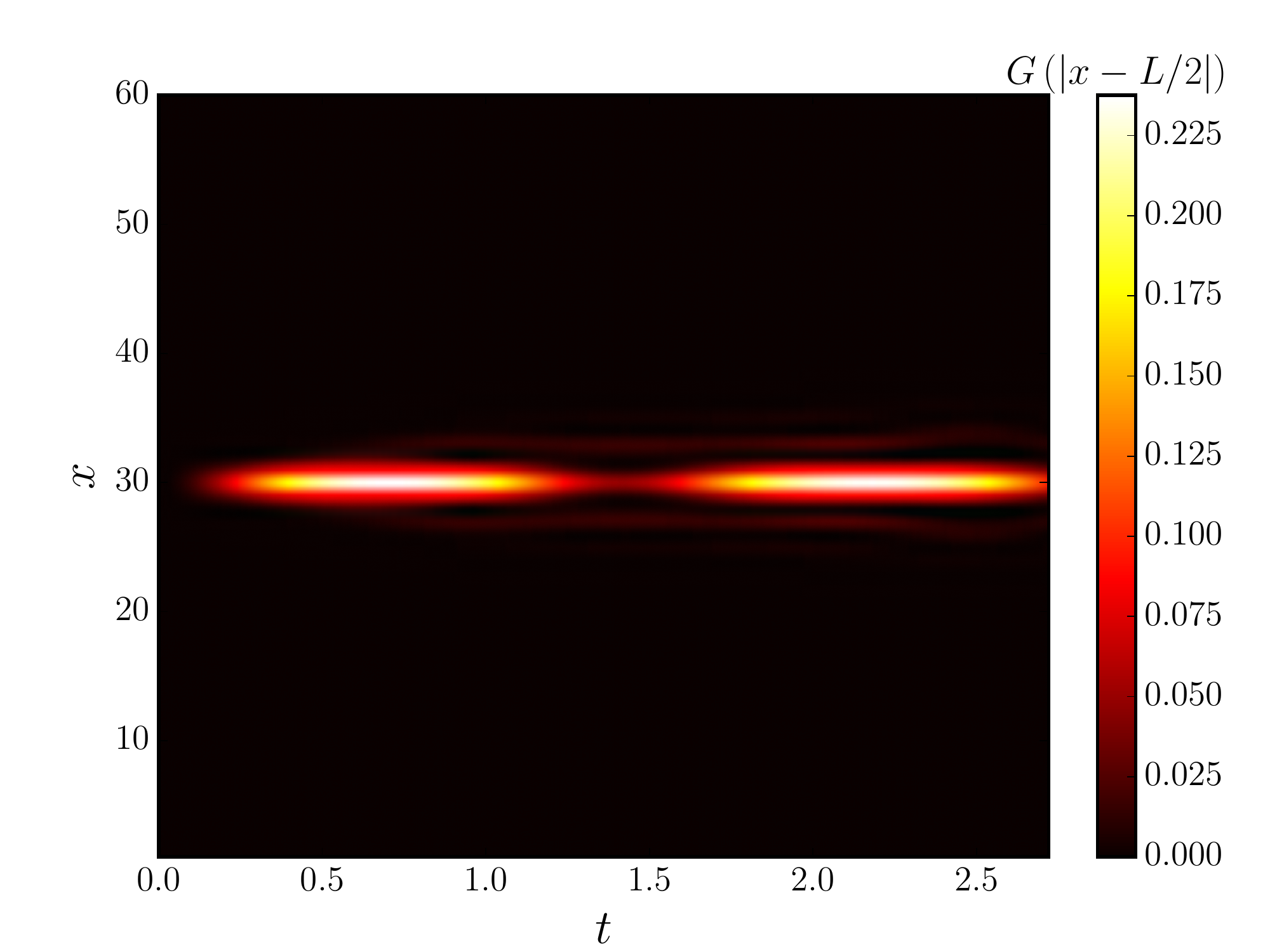}}
\subfigure[\label{fig:g_45}]{\includegraphics[scale=0.4]{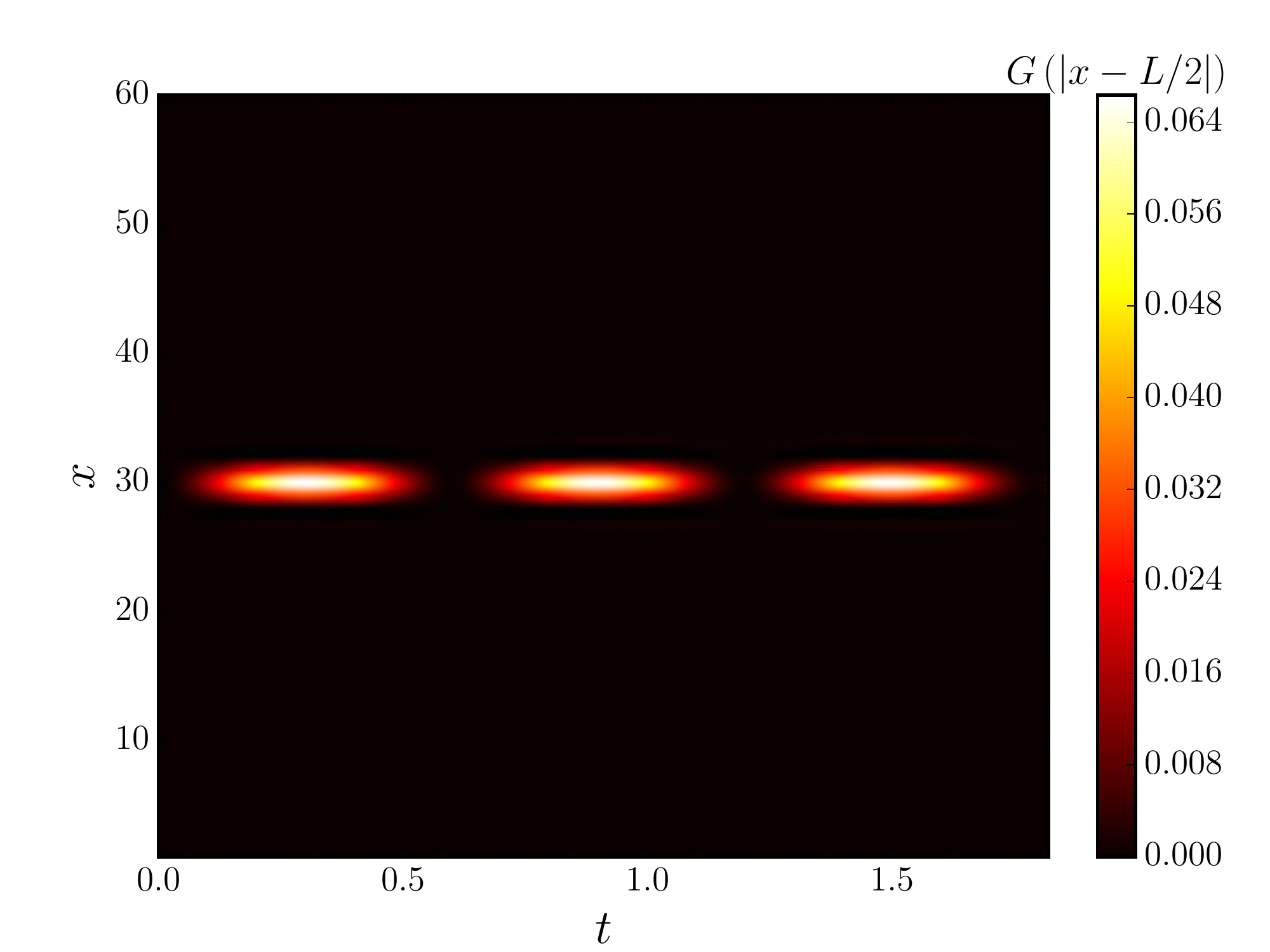}}
\caption{ $\mathbb{Z}_3$-model. Contour plot of the correlation function $G_0(x-L/2)$ of a system initially prepared in the Dirac sea vacuum, for $g=\sqrt{3/\pi}$ and (a) $m=-0.5$ (b) $m=1.0$, (c) $m=4.5$.}
\label{fig:gtime}
\end{figure}

As a last indicator, we consider the time evolution of the connected correlation functions 
\begin{equation}
G_0(x-L/2) (t)= \langle n_{L/2}(t) n_x(t)\rangle - \langle n_{L/2}(t)\rangle \langle  n_x(t)\rangle
\end{equation}
shown in Fig.~\ref{fig:gtime}. For small values of $m$, these functions exhibit a light-cone spreading, typical of conformal or integrable theories, as discussed in~\cite{cc}. However, as we enter in the intermediate and strong confinement region, we observe a localization effect and an oscillatory behaviour, indicating that particle/antiparticles pairs do not spread, but are cyclically created and recombined. Similar behaviors have been observed in other models~\cite{tak,surace}.

\section{Pair production in an external field}
\label{sec:pairfield}

In the context of particle physics, pair production is often studied in presence of an external electric field, from which the pair production rate is expected to depend.
For the Schwinger model, the critical value $E_c$ of the external electric field at which the $e^{+}e^{-}$ production rate should reach a maximum is given by $E_{c}=m^{2}/g$ (which, in SI units, reads $E_c\approx1.32\times10^{18}\,\mathrm{V}/\mathrm{m}$~\cite{Sch51}), but this effect has never been observed experimentally since the aforementioned critical value is still out of the range of even the most powerful lasers. In $1+1$ dimensions, a formula for the production rate has been proposed in~\cite{key-3,key-4}:
\begin{equation}
\dot{\rho}=\frac{eE_{0}}{2\pi}\exp\left(-\frac{\pi m^{2}}{eE_{0}}\right)=\frac{m^{2}}{2\pi}\epsilon\exp\left(-\frac{\pi}{\epsilon}\right)\label{eq:rate_coppie}
\end{equation}
where $\dot{\rho}$ represents the time derivative of the total density of the chain in the infinite length limit, while $\epsilon = E_0/E_c$, $E_0$ being the value of the external field. This formula predicts that the production rate increases linearly with time
for large values of $\epsilon$ and it is exponentially suppressed for $\epsilon\ll 1$.

\begin{figure}[t!]
\subfigure[\label{fig:campo_esterno}]{\includegraphics[scale=0.4]{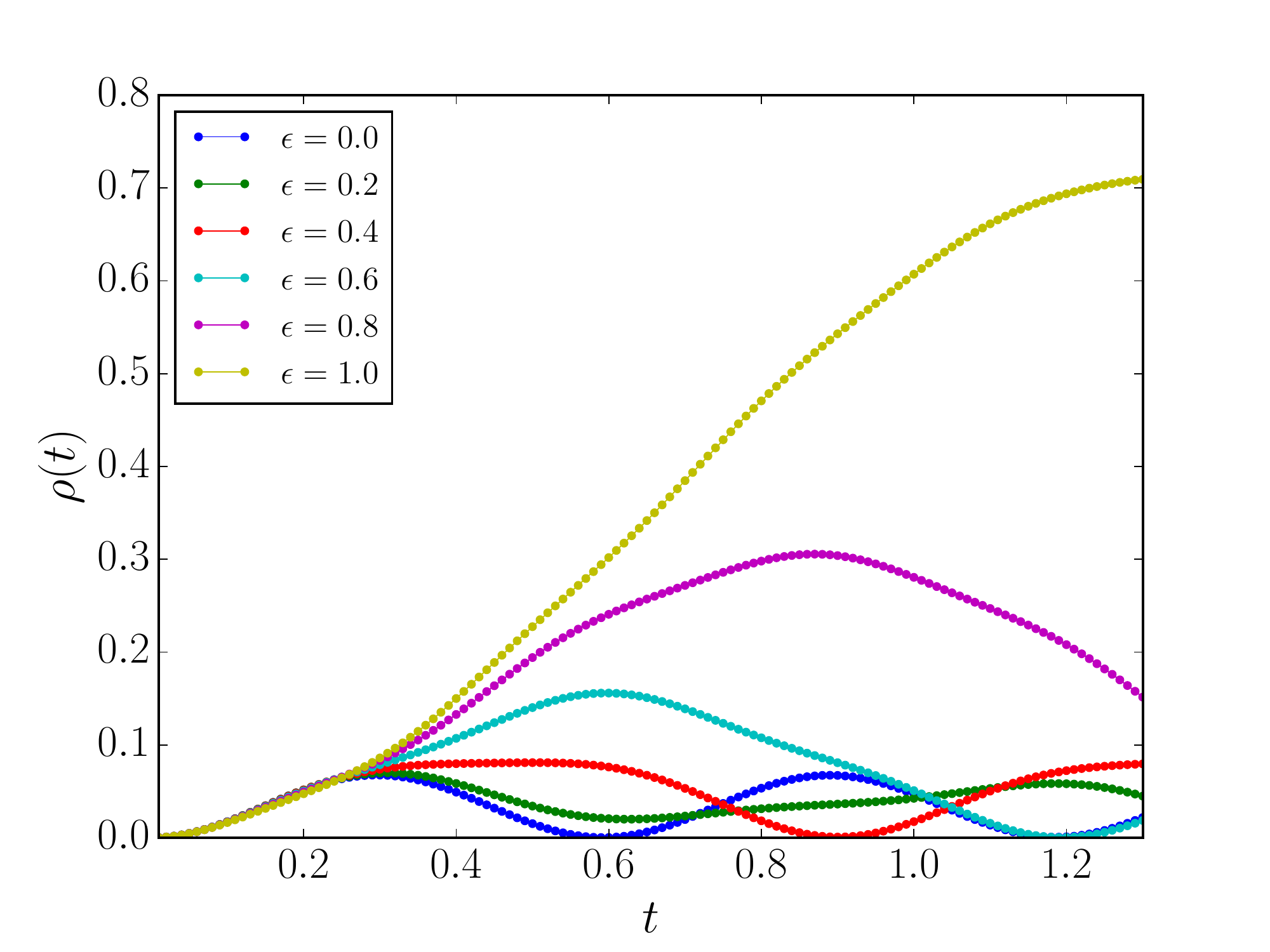}}
\subfigure[\label{fig:fit_lineari_campoesterno}]{\includegraphics[scale=0.4]{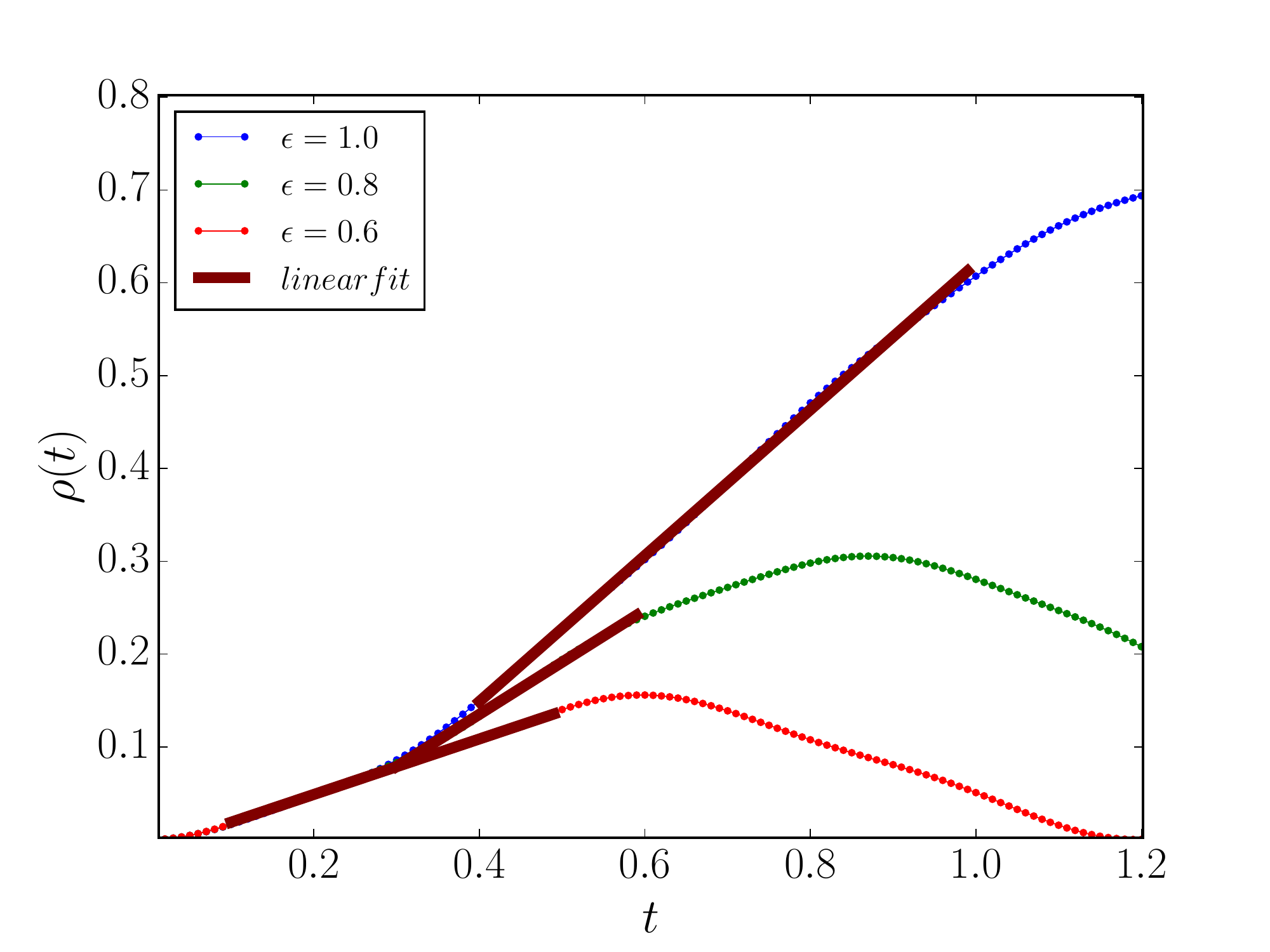}}
\caption{$\mathbb{Z}_3$-model. (a) Particle density $\rho(t)$ of a system initially prepared in the Dirac sea vacuum, represented as function of time for different values of $\epsilon=E_{0}/E_{c}$.
The values of the fixed parameters are $N=50$, $m=4.5$ and $g=\sqrt{3/\pi}$. In panel (b), we report three selected cases, together with linear fits of densities at reasonably small times; these fits are used to evaluate $\dot{\rho}$ and test Eq.~(\ref{eq:rate_coppie}).}
\end{figure}

\begin{figure}
\centering{}\includegraphics[scale=0.4]{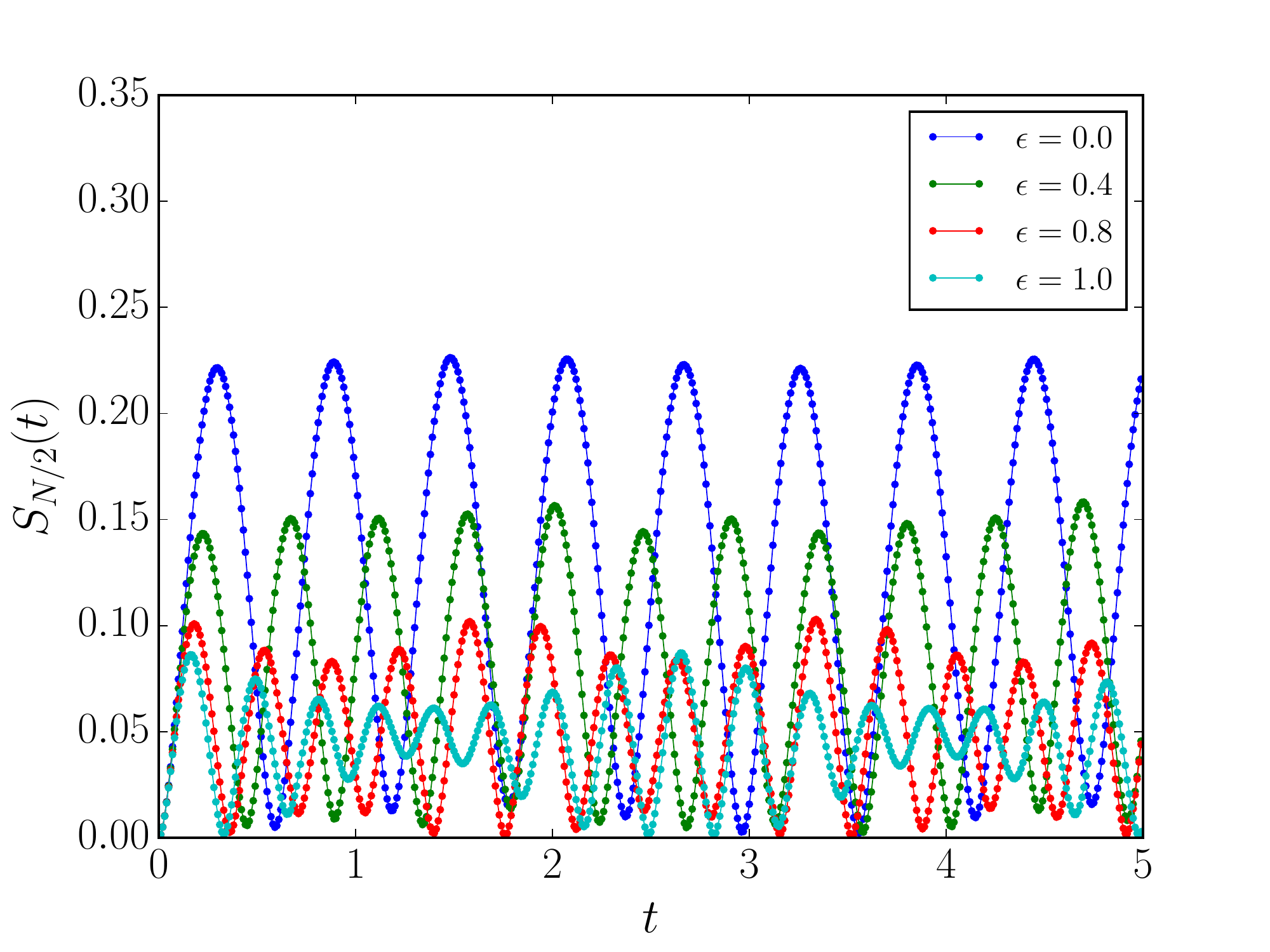}\caption{\label{fig:entropia_esterno} $\mathbb{Z}_3$-model. Half-chain entanglement $S_{N/2}(t)$ of a system initially prepared in the Dirac sea vacuum, evaluated for different values of the scaled external field $\epsilon$. We set $m=4.5$ and $g=\sqrt{3/\pi}$.}
\end{figure}

This formula can be checked in our simulations. We start by considering the $\mathbb Z_3$-model. We choose values of $(m,g)$ for which the Dirac sea vacuum results barely perturbed by spontaneous pair production, according to the analysis described in Sec.~\ref{sec:pair}. At $t=0$, we apply a constant uniform electric field $E_0$ along the whole chain and run the simulation to obtain $\rho(t)$ for the corresponding value of $\epsilon = E_0/E_c$, with $E_c=m^2/g$. Figure~\ref{fig:campo_esterno} shows the results of our simulations for the case $m=4.5$ and $g=\sqrt{3/\pi}$. We observe that the vacuum is stable, with some small oscillations, for very small values of $\epsilon$. As we increase $\epsilon$, we start observing a linear regime, as predicted by Eq.~(\ref{eq:rate_coppie}), followed by a saturation effect. Figure~\ref{fig:fit_lineari_campoesterno} shows the range in which we performed the linear fit in order to evaluate $\dot{\rho}$ and verify Eq.~(\ref{eq:rate_coppie}). 

Even if we do predict a linear regime in accordance with the continuous case, our simulations suffer from non negligible finite-size and small-$n$ effects, as discussed in the Appendix. There, we plot data for different values of $N$ (namely, $N=50,60,70,80,90$) as well as for the $\mathbb Z_5$ and $\mathbb Z_7$ models. Our results show that the approximation of the continuum limit improves as we increase $N$ and $n$.

Finally, in order to get additional insights, we calculated the time evolution of the half-chain entropy, for different values of the external field. A typical behavior is shown in Fig.~\ref{fig:entropia_esterno}, which displays the data for $m=4.5$ and $g=\sqrt{3/\pi}$. We observe that the entropy always presents an oscillatory behavior, with not-too-large maximum values. More interestingly, we observe that an increase of the external field, which results in a rapid increase in particle pair production, as shown in Fig.~\ref{fig:campo_esterno}, does not contribute at all to the increase and spread of entanglement. 
It seems that the mechanism that dictates the dynamics of the particle/antiparticle pair production is different in absence and in presence of the external field: in the former case, entangled quark/antiquark pairs are spontaneously created out of the Dirac sea and spread along the chain, at a mass-dependent speed. On the contrary, in the latter situation, the electric charges rearranges only locally to form mesonic couples that allow for minimization of the interaction with the external electric field, without any spreading.

\section{The string breaking mechanism}
\label{sec:string}

As recalled in the Introduction, the string breaking mechanism is a very interesting phenomenon that is expected to occur in theories, such as $(3+1)$-QCD, admitting confined phases, but that is also very hard to prove, either analytically or numerically, since its effects are mainly dynamical. 
In the framework described in this article this phenomenon can be investigated, since one can simulate the real-time dynamics of a generic initial state, which is let to evolve with the discretized Hamiltonian (\ref{HamQED2}). In particular, the system is prepared in the string-excitation state shown in the last configuration of Fig.~\ref{fig:2}, where a particle and an antiparticle separated by a distance $l$ are created from the Dirac sea vacuum, giving rise to a non-zero string of electric field in between. Here, we want to explore the effects on dynamics of both coefficients appearing in the Hamiltonian: therefore, we perform the numerical simulations for different values of $m$ and $g$.

\begin{figure}
\hspace{0.3cm}\subfigure[\label{fig:sb1v}]{\includegraphics[scale=0.4]{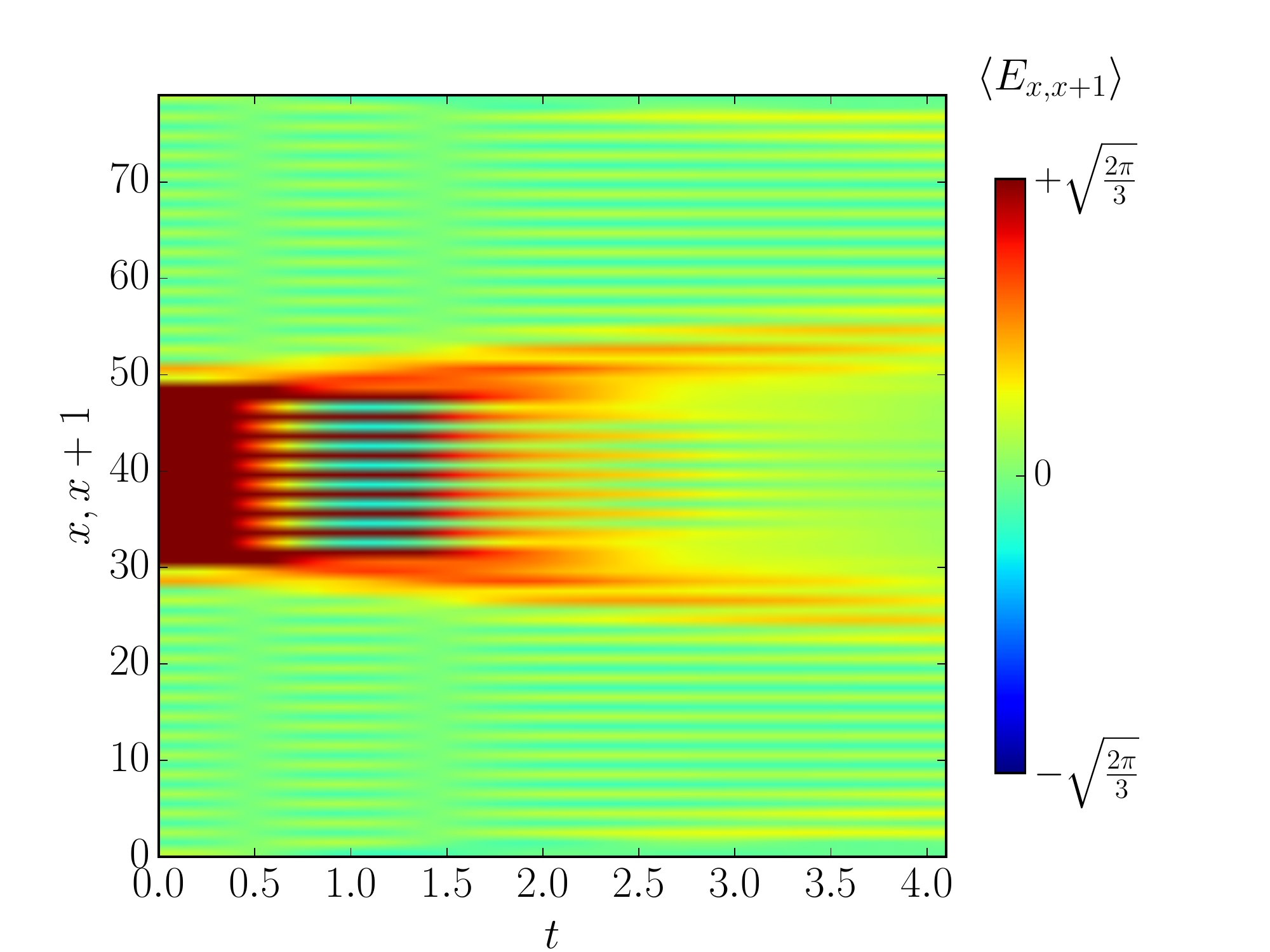}}
\subfigure[\label{fig:sb2v}]{\includegraphics[scale=0.29]{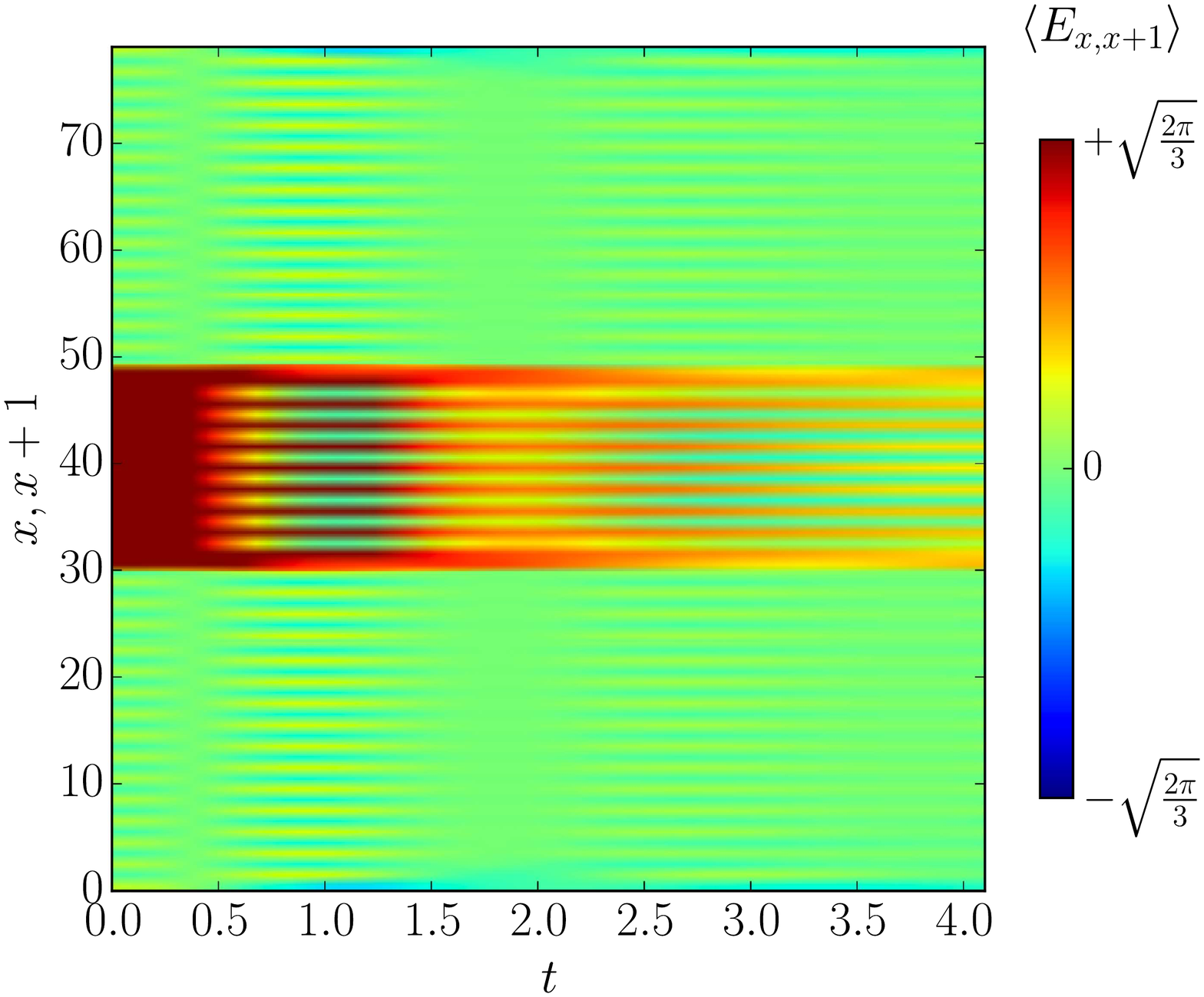}}
\subfigure[\label{fig:sb3v}]{\includegraphics[scale=0.29]{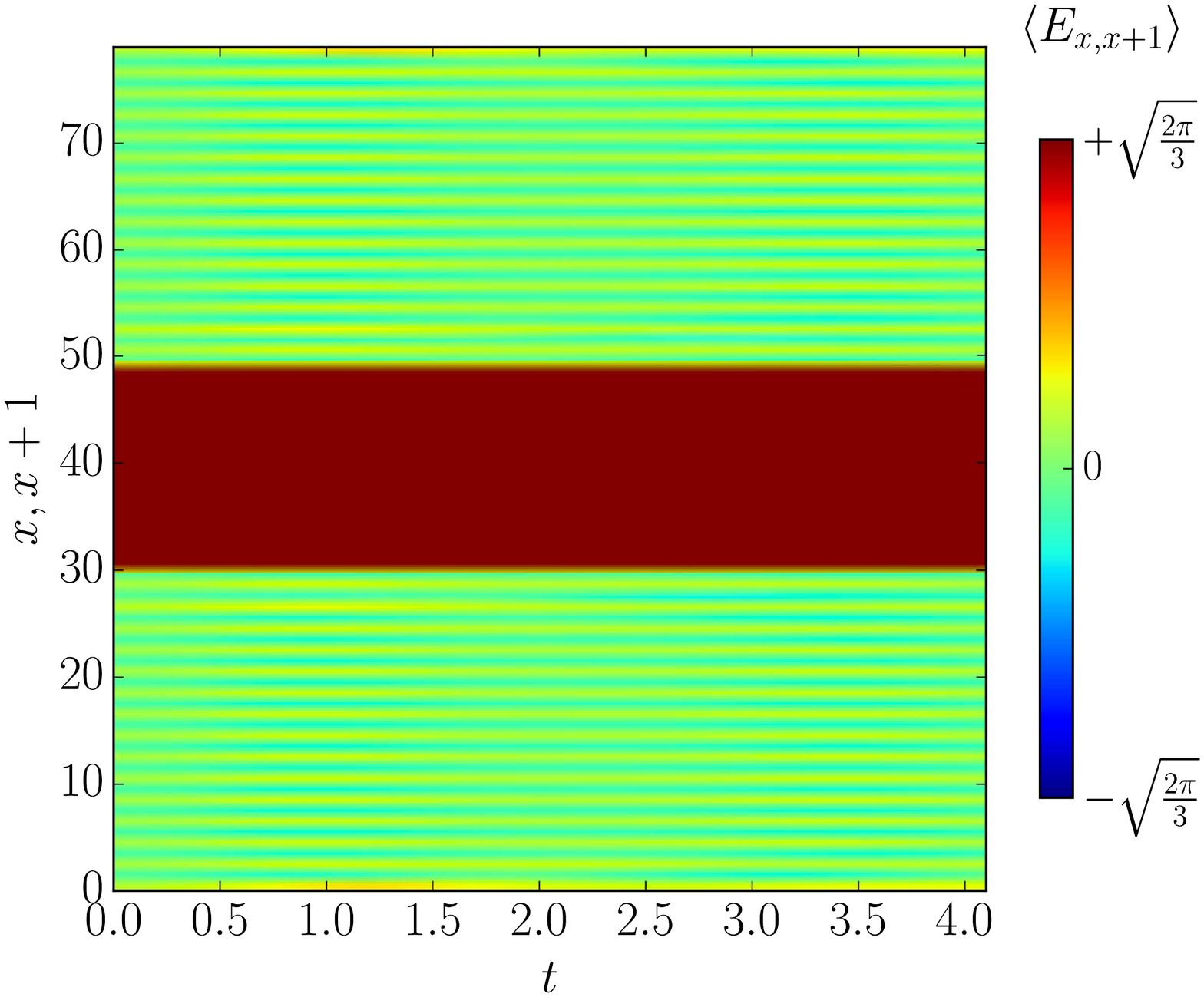}}
\caption{$\mathbb{Z}_3$-model. Real-time dynamics of a string: evolution of the electric field $E_{x,x+1}(t)$ on links, for (a) $(m=0.1,g=0.1)$, (b) $(m=0.3,g=0.8)$, (c) $(m=3.0, g=1.42)$. As explained in the text, the plotted value of the electric field is that obtained after subtracting the effects due to spontaneous pair production, which tends to blur the dynamics of the string. }
\label{fig:sbv}
\end{figure}

In the following, we will consider the $\mathbb Z_3$ model and a chain of length $N=80$. We initialize our system in a state with a string placed at the center of the chain: the particle and the antiparticle are put at a distance equal to 20 lattice sites, so that the electric field is different from zero (and equal to $+\sqrt{2\pi/3}$) only on the $19$ central links of the chain. The evolution of this state is followed by looking at the value of the electric field $E_{x,x+1}(t)$ on each link. We report the analysis of the real time dynamics of such a string for three different values of $(m,g)$: (a) $m=0.1$, $g=0.1$, (b) $m=0.3$, $g=0.8$, (c) $m=3.0$, $g=1.4$. The total evolution of the initial state is also affected by the spontaneous pair production that, as we examined in the previous section, gives rise to a local non-zero field. Thus, to isolate the additional effects due to string breaking and to improve the interpretation of our numerical results, we subtract from $E_{x,x+1}(t)$ the value of the electric field that would be obtained in the evolution of  Dirac sea vacuum for the same values of $(m,g)$. The data, corrected accordingly, are shown in Fig.~\ref{fig:sbv}. They clearly show three different situations:
\begin{itemize}
\item[a)] in Fig.~\ref{fig:sb1v}, the string starts to spread and breaks into particle/antiparticle pairs (mesons) that, after a short time during which one can notice a rich process of pair production and recombination, stabilize in a configuration with two mesons localized at the edges of the string; the two mesons are deconfined, since they move away one from each other at a constant speed. A rough estimate of this speed can be obtained by estimating the meson ``positions'' at the end of the simulation from the peak values of the electric field, and then measuring the distance each meson has traveled during the time of the simulation; the speed decreases as either of the parameters is increased, approaching zero as we enter the region described in the next case;
\item[b)] in Fig.~\ref{fig:sb2v}, the string does not spread; still, it breaks into particle/antiparticle pairs (mesons) that, as in the previous case, quickly stabilize in a configuration where we can still distinguish two mesons localized at the edges of the string; however, now the two mesons are confined, since they remain at a fixed distance from each other;
\item[c)] in Fig.~\ref{fig:sb3v}, the string is completely stable: it does not spread and it does not break into mesons. 
\end{itemize}

\begin{figure}
\begin{centering}
\includegraphics[scale=0.47]{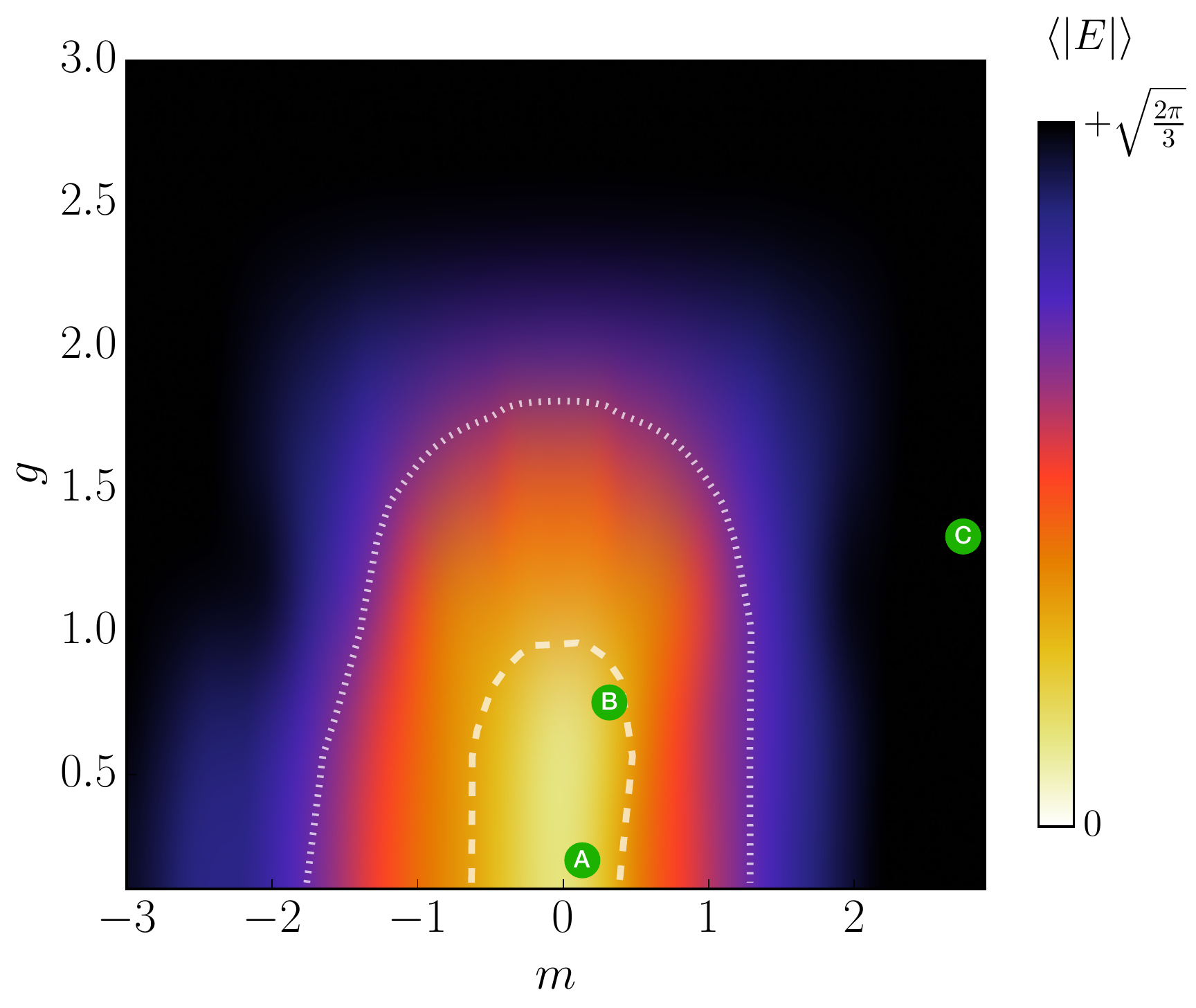} \caption{\label{fig:string_breaking_diagramma} Contour plot in the $(m,g)$ plane of the asymptotic total value of the electric field at the center of the chain. The two white level curves correspond to $10\%$ (dotted line) and $50\%$ (solid line) of the initial value, respectively.
The three points $A,B,C$ correspond to the values of the parameters $(m,g)$ chosen for the three simulations shown in Fig.~\ref{fig:sbv}.}
\par\end{centering}
\end{figure}

The simulations shown in Fig.~\ref{fig:sbv} can be repeated for any couple of values $(m,g)$. Our results are summarized in Fig.~\ref{fig:string_breaking_diagramma}, where we show the contour plot of the large-$t$ (namely $t=4.0$ in our units) total value of the electric field at the center of the chain, defined as the sum of the electric field on the 12 central links, as a function of the coupling constants $(m,g)$. The two (white) level curves correspond to $10\%$ (dotted line) and $50\%$ (solid line) of the initial value, respectively. From this picture, the three regimes described above are clearly identified: (a) a weak confinement regime in the lighter and central part of the diagram, corresponding to the breaking of the string into two deconfined mesons; (b) an intermediate confinement regime in the reddish part of the diagram, corresponding to the breaking of the string into two confined mesons; (c) a strong confinement regime in the darker part, corresponding to a stable string configuration. 

\begin{figure}
\subfigure[\label{fig:sbe1}]{\includegraphics[scale=0.4]{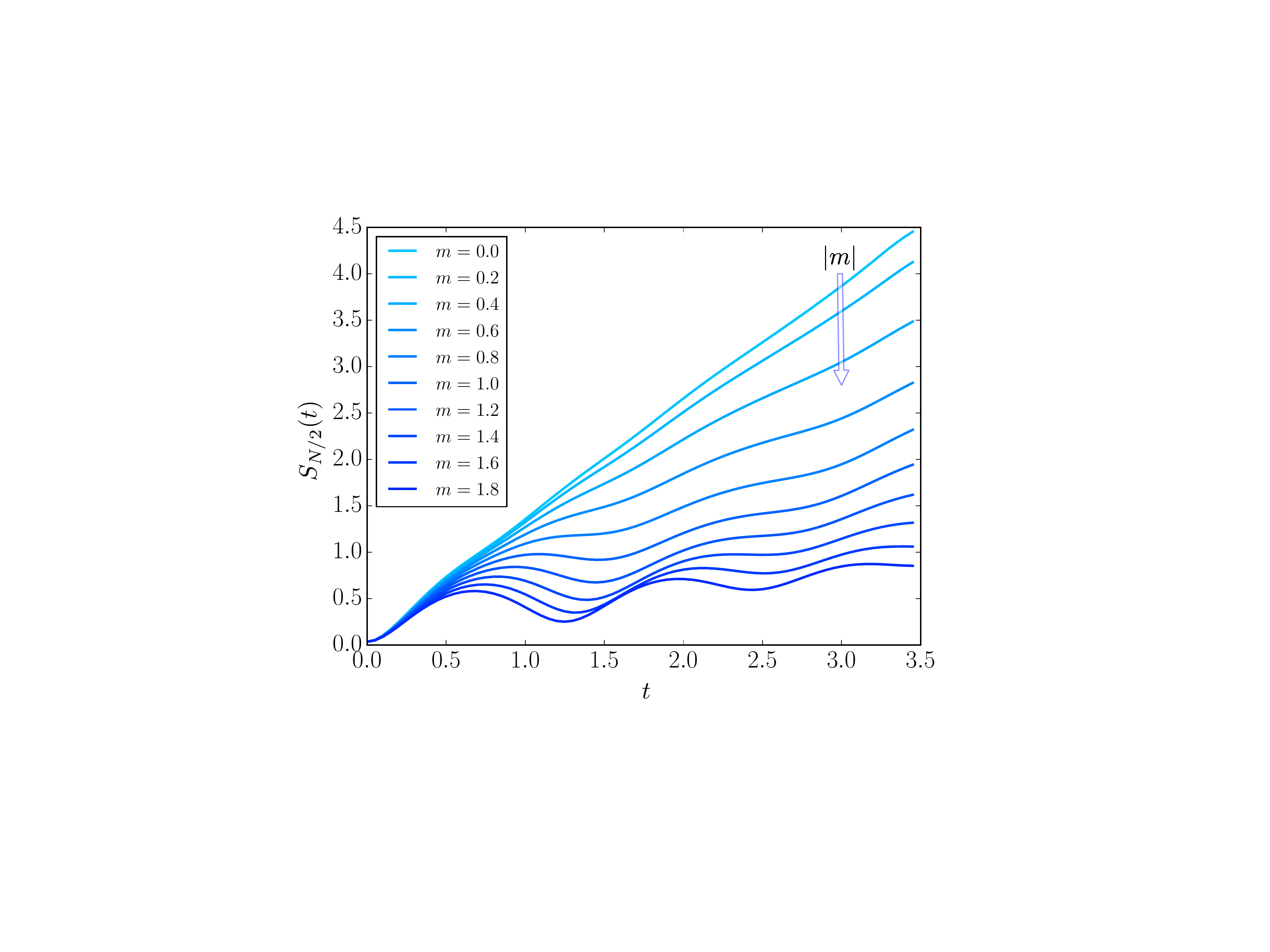}}
\subfigure[\label{fig:sbe2}]{\includegraphics[scale=0.4]{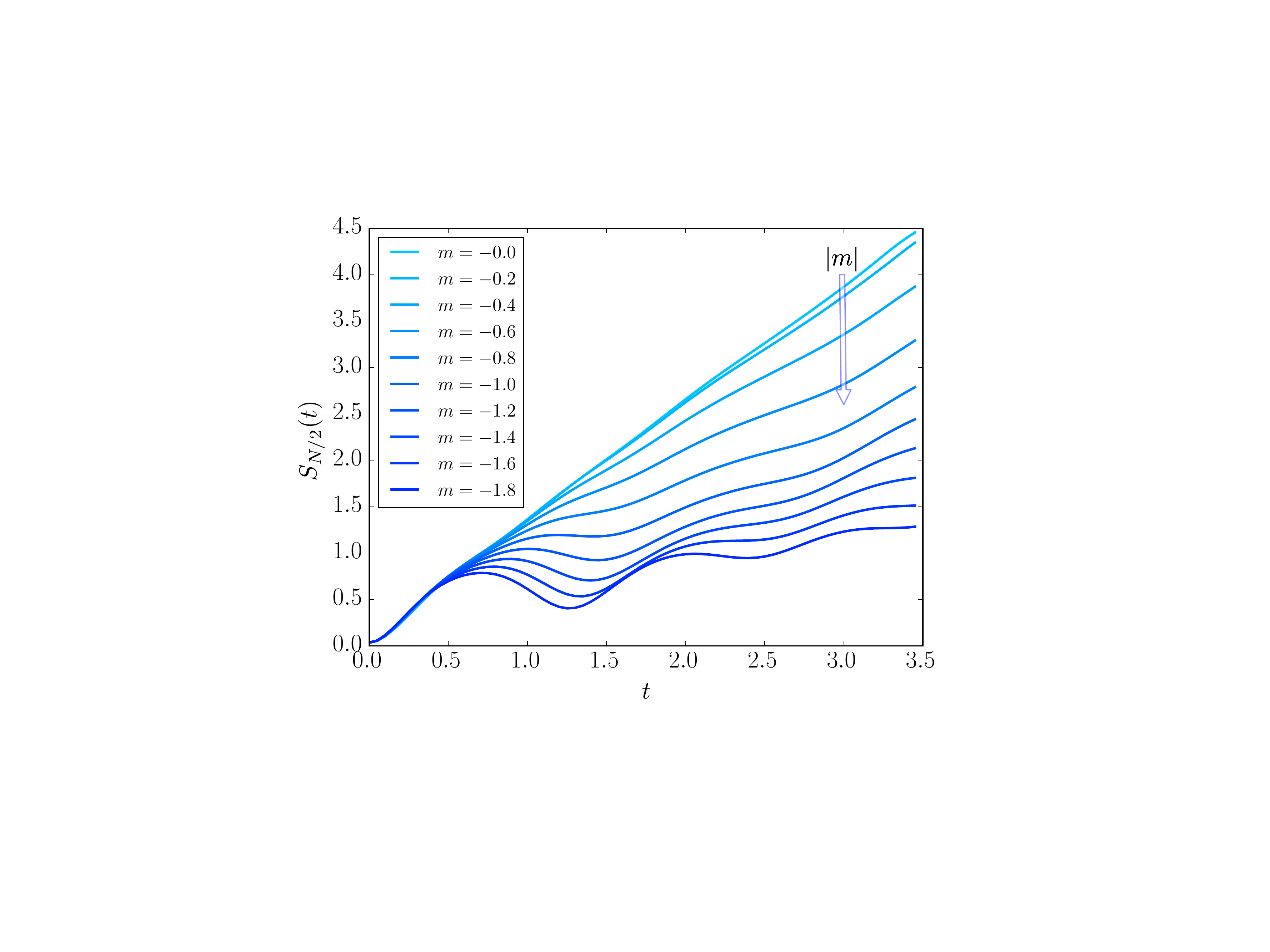}}
\subfigure[\label{fig:sbe3}]{\includegraphics[scale=0.4]{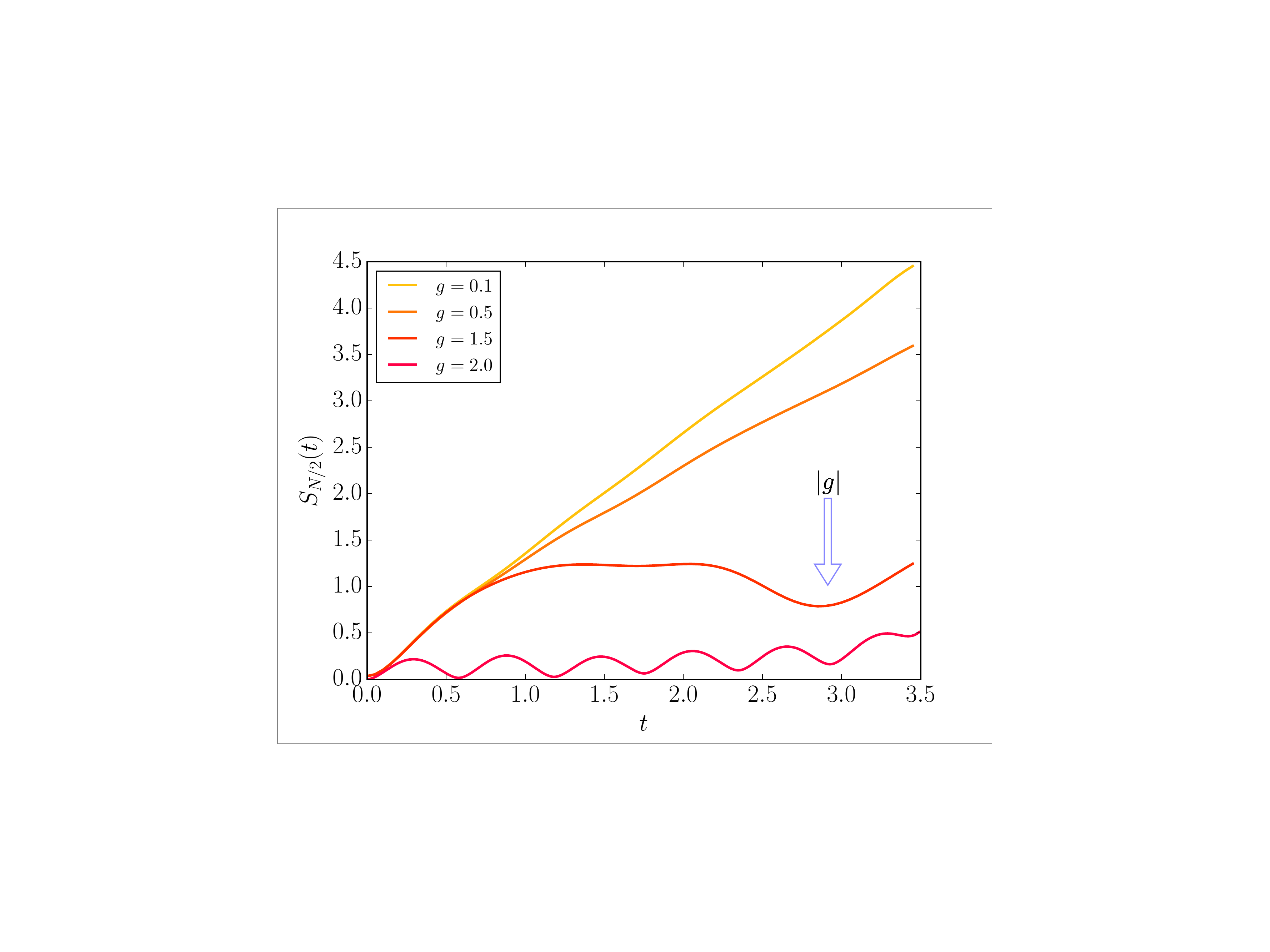}}
\caption{(a) $\mathbb{Z}_3$-model. Time evolution of the half-chain entropy $S_{N/2}(t)$ for: (a) $g=0.1$ and different positive values of $m$,  (b) $g=0.1$ and different negative values of $m$, (c) $m=0.1$ and different values of $g$.}
\label{fig:sbe} 
\end{figure}

It is interesting to examine the string breaking phenomenon by looking also at the time evolution of the half-chain entanglement entropy, for different values of $(m,g)$. Figure~\ref{fig:sbe} shows the behaviour of $S_{N/2}$ evaluated on the time evolution of the string state, by first keeping $g=0.1$ fixed and letting $m$ vary with (a) positive, and (b) negative values; and then by (c) keeping $m=0.1$ fixed and letting $g$ vary. These graphs confirm and corroborate what was found by looking at the real time dynamics of the electric field configuration: namely a growth of the entanglement entropy that is linear for small $m$ and/or small $g$, sublinear for intermediate values, and suppressed (but for small oscillations) in the strong confinement regime. 

\section{Conclusions}
\label{sec:concl}

We have investigated the out-of-equilibrium properties of $(1+1)$-dimensional QED, approximated via a $\mathbb{Z}_{n}$ Schwinger model. By means of simulations, focusing on the stability of the Dirac vacuum with respect to particle/antiparticle pair production and on the string breaking mechanism, we have studied the effects of confinement on the real-time dynamics of the model.
We have found that confinement is not a feature of the $U(1)$ Schwinger model only, having a relevant effect on the dynamical properties of the $\mathbb{Z}_{n}$-model as well, as it is proved by the enhanced oscillatory behaviour of the relevant physical observables in the process of pair production, and by the total suppression of the breaking and spreading of string excitations, with a perfect localization of the latter.

Let us notice that such a reduction of entanglement and slow-down of the dynamics have been observed in other systems. This is the case not only of models with long-range interactions~\cite{pagano}, but also of the Ising and Potts models with both a transverse and a longitudinal magnetic fields~\cite{tak,octavio}. A similar behaviour has also been seen in constrained models which exhibit quantum scar states~\cite{scar1,scar2} and in spin-$1/2$ chain Hamiltonians that can be derived from Abelian gauge models in the quantum link approach~\cite{surace}. It is interesting to notice that, in the two latter cases, the physical states of the system under consideration are constrained to lie in a restricted subspace of the total Hilbert space, a fact that is shared by our model, where the role of Gauss's law constraint is crucial. This is also at the heart of the recent proposal to experimentally implement these Hamiltonians with Rydberg atomic systems in the Rydberg-blockade regime~\cite{simone}. We plan to further investigate the role of the gauge constraint in future work.

\section*{Acknowledgments} The authors are grateful to Fabio Ortolani for his precious help with the DMRG code.
EE and GM are are partially supported by Unibo through the project ``ALMAIDEA''.
FVP is supported by INFN through the project ``PICS''.
PF is partially supported by the Italian National Group of Mathematical Physics (GNFM-INdAM). MD is partly supported by the QUANTERA project QTFLAG, and by the ERC under grant number 758329 (AGEnTh).
EE, PF, GM, FP and SP are partially supported by INFN through the project ``QUANTUM'' and QuantERA ERA-NET Cofund in Quantum Technologies (GA No. 731473), project QuantHEP. 
PF, FP and SP are partially supported by Regione Puglia.

\appendix
\section{Finite size scaling and large-$n$ limit}
\label{addinfo}

\begin{figure}
\centering
\subfigure[\label{fig:scaling_massimo}]{\includegraphics[width=0.48\textwidth]{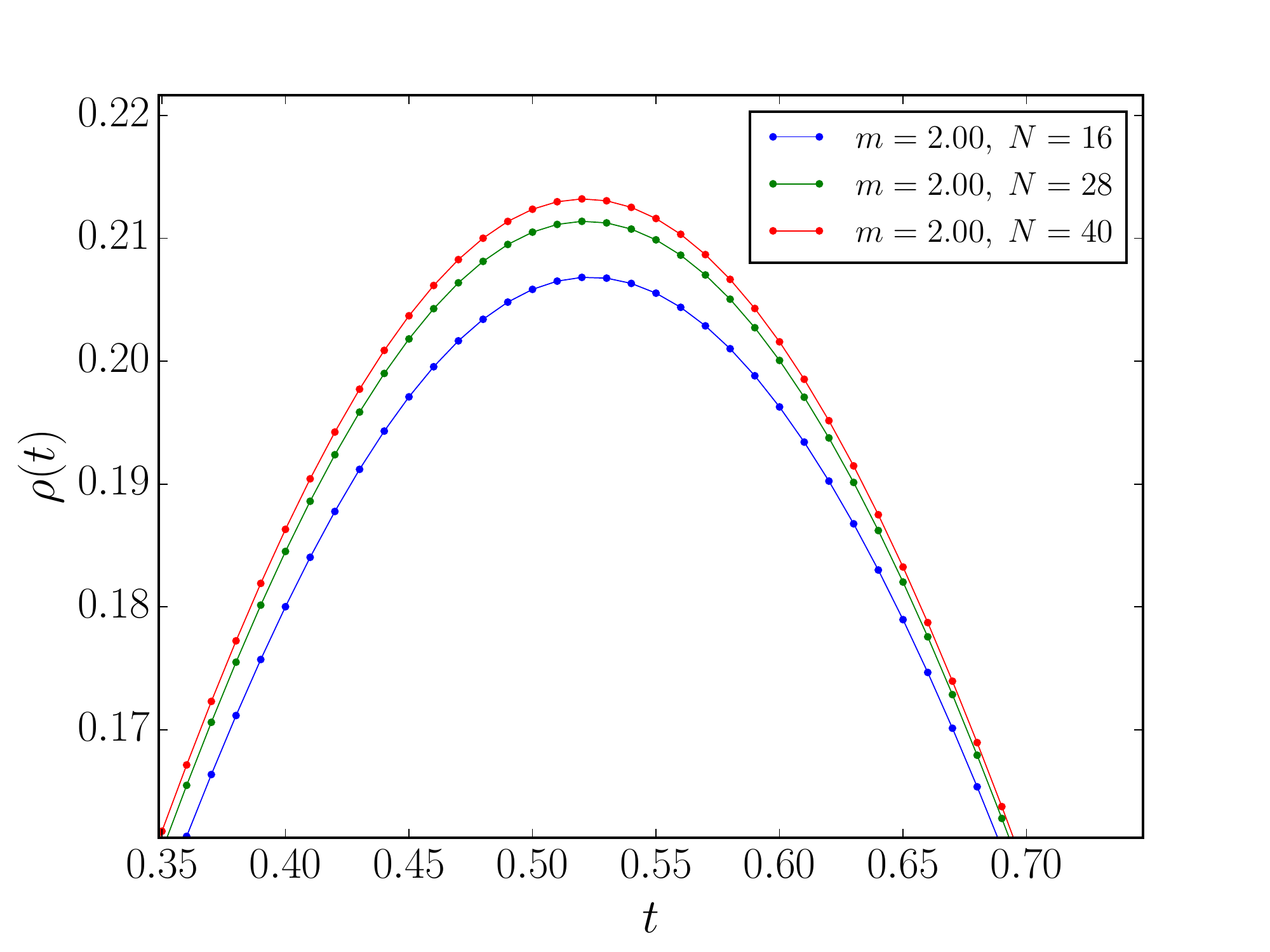}}
\subfigure[\label{fig:fit_scaling1}]{\includegraphics[width=0.48\textwidth]{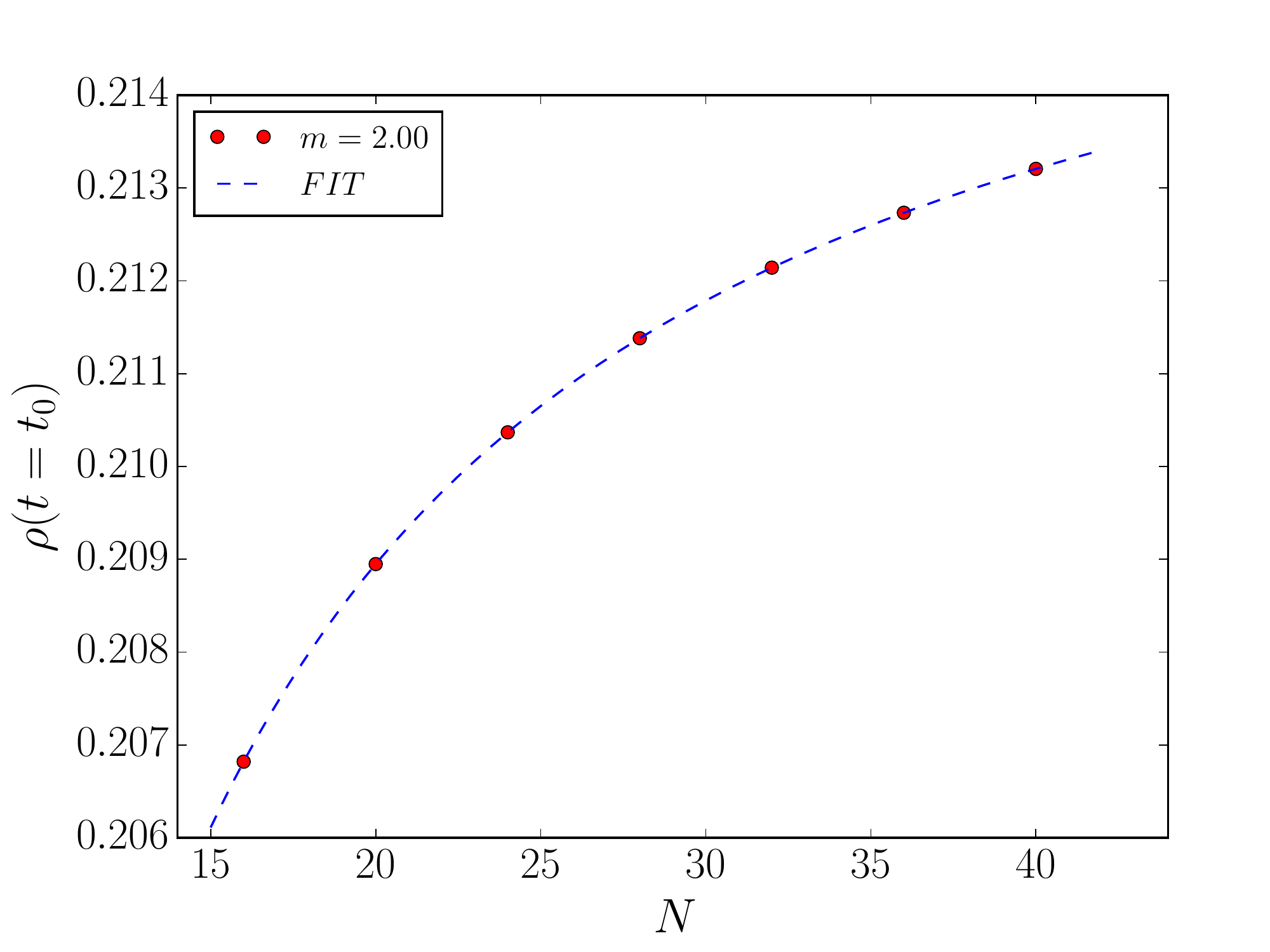}}
\subfigure[\label{fig:densita_estrapolata_m_fissato}]{\includegraphics[width=0.42\textwidth]{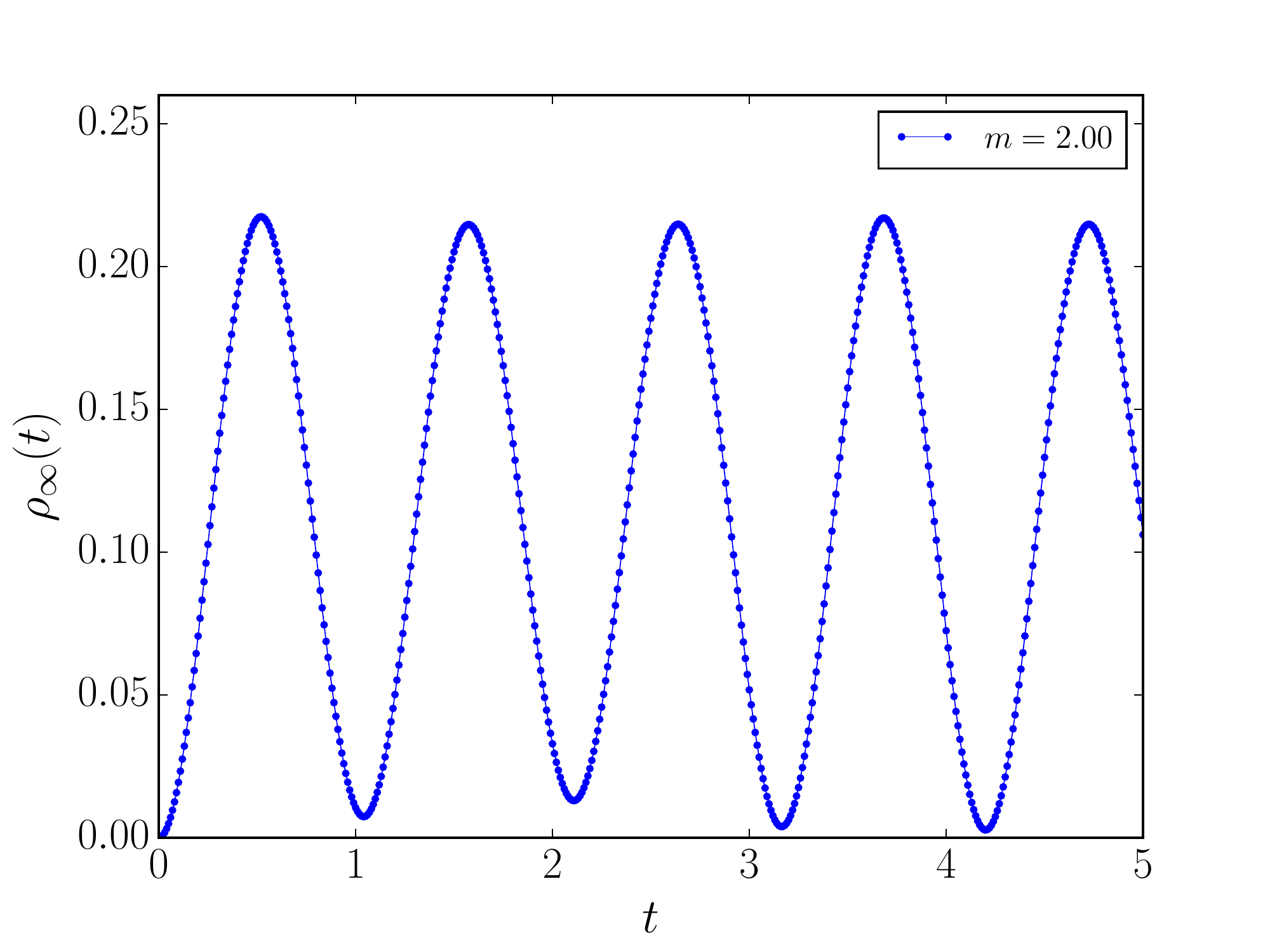}}
\caption{$\mathbb{Z}_3$-model.  Finite size analysis of the particle density, for $m=2.0$; (a) behaviour of $\rho(t)$ close to its first maximum;  (b) scaling of $\rho(t_0=0.52)$ with the chain size $N$; (c) time evolution of $\rho_{\infty}(t)$. \label{fig:scaling_m}}
\end{figure}

\begin{figure}
\centering
\subfigure[\label{fig:densita_estrapolata_z5}]{\includegraphics[width=0.48\textwidth]{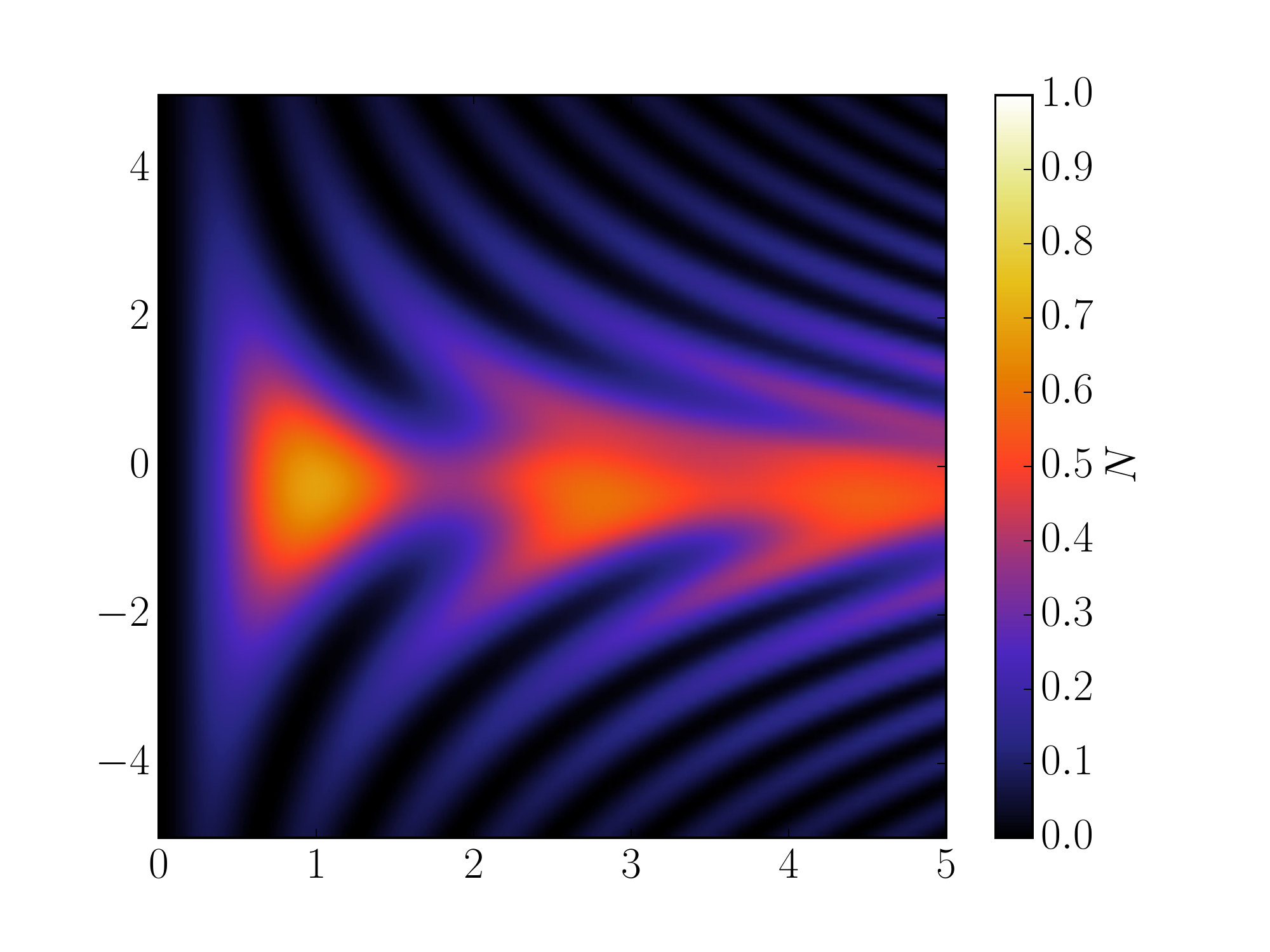}}
\subfigure[\label{fig:densita_estrapolata_z7}]{\includegraphics[width=0.48\textwidth]{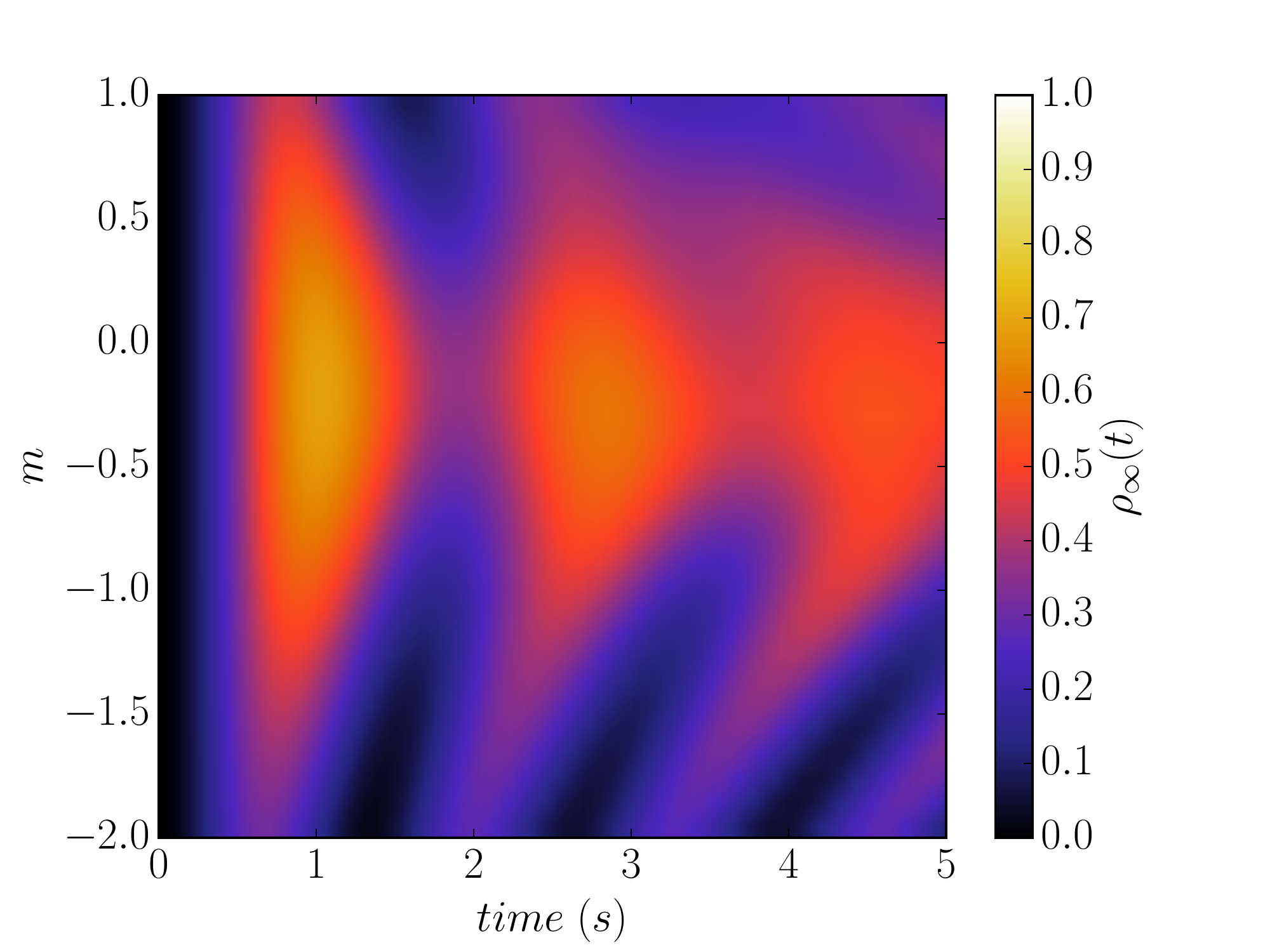}}
\caption{Contour plot of $\rho_\infty (t)$ for (a) $\mathbb{Z}_5$-model and $m\in[-5.0,+5.0]$; (b) $\mathbb{Z}_7$-model and $m\in[-2.0,+1.0]$. Notice that in the latter case we reduced the mass range, since numerical simulations are computationally cumbersome.}
\end{figure}

The question about how our lattice $\mathbb{Z}_n$ model converges to the $U(1)$ continuum model was theoretically studied in~\cite{NEFMPP} and then thoroughly checked via numerical simulations in~\cite{Erc}. There we noticed that one recovers the Schwinger model for QED in $1+1$ dimensions, obtaining a good approximation already for lattice sizes of the order of $50$ sites and $n=3$.  We refer the interested reader to these references for details on how one can rigorously control both the continuum and the finite-$n$ limits. In this Appendix we give some additional details on the numerics and finite-size and large-$n$ analyses we  performed for the time-dependent simulations.

\begin{figure}
\centering
\subfigure[\label{fig:confronto_con_rate1}]{\includegraphics[width=0.48\textwidth]{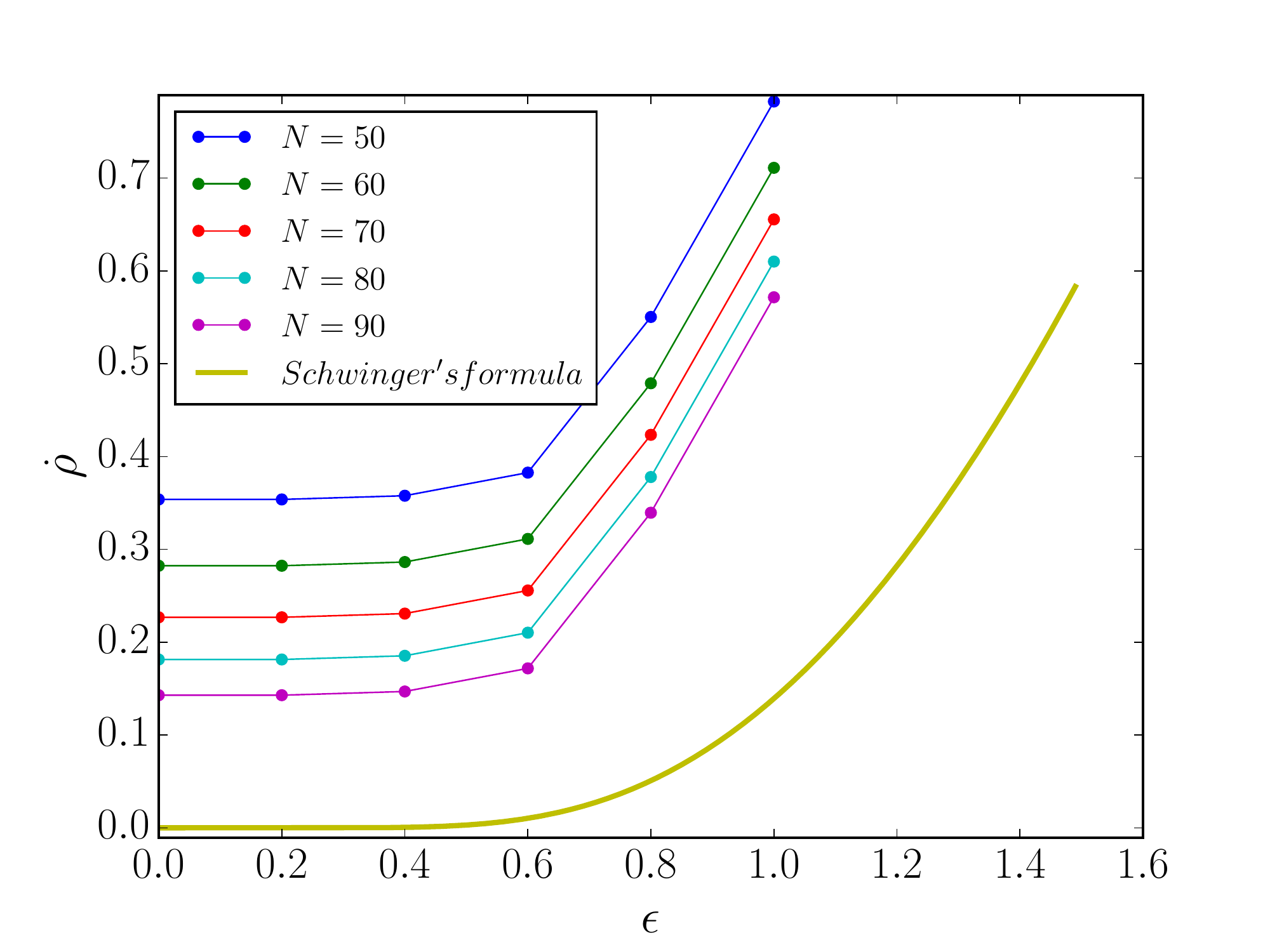}}
\subfigure[\label{fig:confronto_con_rate2}]{\includegraphics[width=0.48\textwidth]{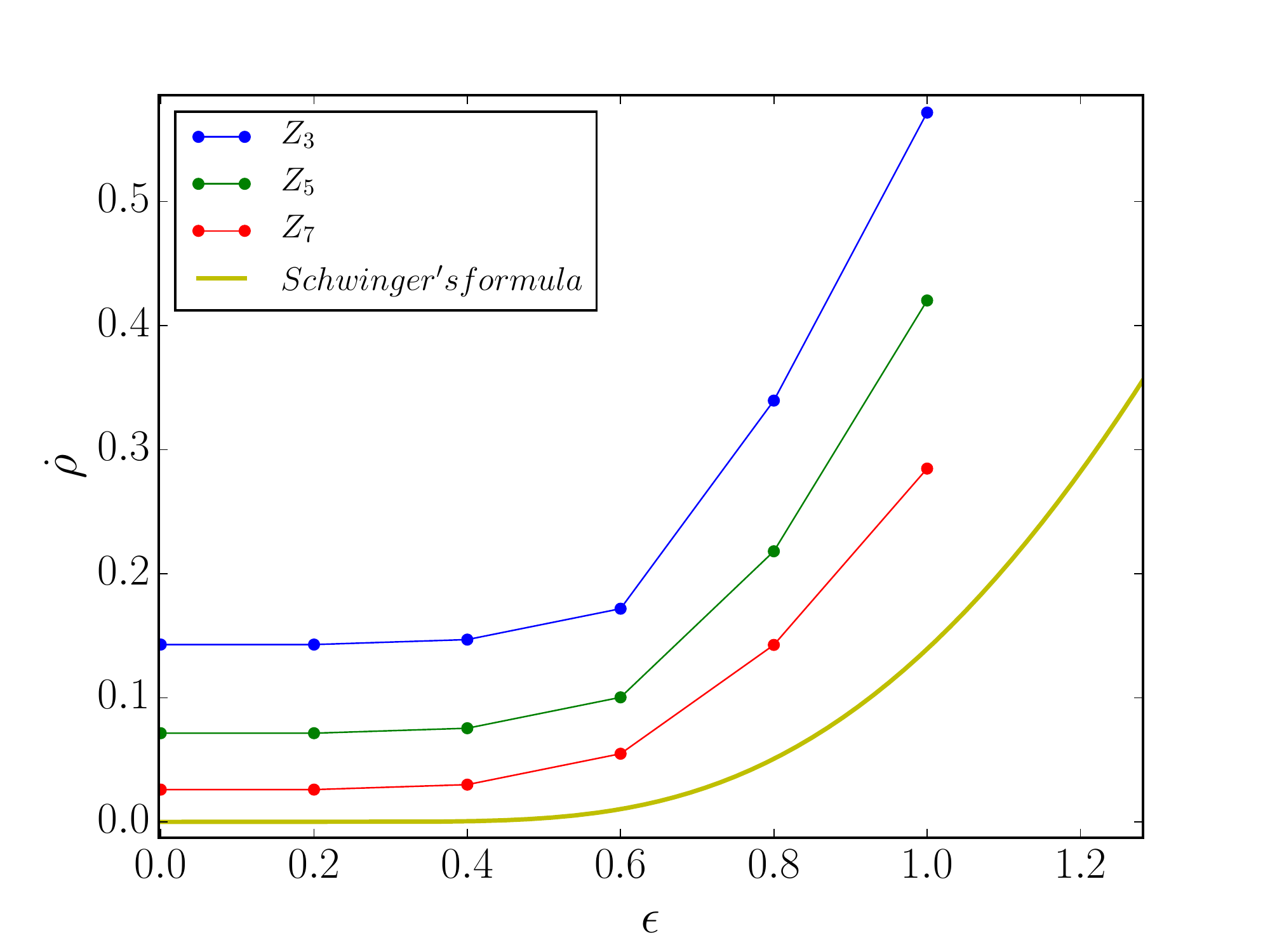}}
\caption{(a) Pair production rate $\dot{\rho}$ in the $\mathbb{Z}_3$ model as function of $\epsilon$, for different chain size $N$; the continuous result of Eq.~(\ref{eq:rate_coppie}) is shown for comparison. (b) Pair production rate $\dot{\rho}$ as function of $\epsilon$ for different $\mathbb Z_n$ models (at fixed $N=90$)  and comparison with the continuous result of Eq.~(\ref{eq:rate_coppie}). }
\end{figure}

\subsection{Numerical precision} 
All simulations have been performed with a time-dependent DMRG (t-DMRG) code, which is a well-established method for studying the dynamical properties of quantum systems in one dimension~\cite{t_DMRG}. The time-evolution is based on a Runge-Kutta 4th order scheme, with a time step of $\delta=0.01$. This allows us to reach a time of the order of $t_{max} \sim 4-5$ by keeping a unitarity evolution with a precision in the norm of at least $93\%$ of the initial value.

 The initial state of the time-evolution is implemented by calculating the ground-state of two different Hamiltonians: i) for the pair production analysis,  the vacuum state (see Fig.~\ref{fig:2}) is obtained by using the Hamiltonian (\ref{HamQED2}), setting $t=0$ and choosing large values of $m$ and $g$ (i.e.\ $m=5$, $g=3$); ii) for the string breaking mechanism, we added to the aforementioned Hamiltonian localized electric field terms $E_{x,x+1}$ on each link in which the initial string is created. In our simulations, we used a variable number of DMRG-states, up to 1200, in order to keep the truncation error below $10^{-6}$ at each time step.

\subsection{Spontaneous pair production} To study finite size effects, we repeated our simulations for different chain size ($N=16,20,24,28,32,36,40$), in order to extract the infinite size limit of $\rho(t)$. For example, Fig.~\ref{fig:scaling_massimo} shows the behaviour of $\rho(t)$ close to its first maximum, for $m=2.0$ and different chain sizes. We find that, at all instants of time $t_0$, the infinite size limit $\rho_\infty(t_0)$ can be extrapolated according to the following fit:
\begin{equation}
\rho(t=t_{0})=\rho_{\infty}(t_{0})-\frac{\beta(t_{0})}{N}.\label{eq:scaling_densita}
\end{equation}
An example, for $m=2.0$ and $t_0=0.52$ is shown in Fig.~\ref{fig:fit_scaling1}, for which we get $\rho_{\infty}(t_{0})=0.2175\pm0.0001$ and 
$\beta(t_{0})=0.1703\pm0.0001$. The time evolution of $\rho_{\infty}(t)$ is given in Fig.~\ref{fig:densita_estrapolata_m_fissato}. 

We also repeated the same simulations for $\mathbb Z_5$ and $\mathbb Z_7$. In Figs.~\ref{fig:densita_estrapolata_z5}
and~\ref{fig:densita_estrapolata_z7} we show the contour of $\rho_\infty (t)$ in the whole range  of the quenched mass $m\in[-5,+5]$, similarly to what was done in Fig.~\ref {fig:densita_estrapolata_ogni_m} for $\mathbb Z_3$. A very similar behavior is observed. \\

\subsection{Pair production in an external field}
To evaluate finite-size effects in the pair production rate, we repeated the simulations for $N=50,60,70,80,90$. In order to check the large-$n$ limit and verify that the $U(1)$ limit can be reasonably approximated, we performed simulations for the $\mathbb Z_5$ and $\mathbb Z_7$ models. The results are shown and compared with the Schwinger formula in Fig.~\ref{fig:confronto_con_rate1} and Fig.~\ref{fig:confronto_con_rate2}. Even at the largest system sizes that can be numerically investigated, it is evident that the numerics still suffers from significant finite-size and finite-$n$ corrections. However, the data follow a pattern that is qualitatively in agreement with the predictions of the continuum $U(1)$-model and show the correct scaling. We conclude that the system describes the physics of the Schwinger model.


\begin{thebibliography}{10}
\bibitem{f1}
Compare for example our Fig.~\ref{fig:entropia_m_vicini_a_zero} with Fig.\ 6 of Ref.~\cite{tak}.
\end{thebibliography}
\end{document}